\documentclass[a4paper,traditabstract,longauth]{aa}

\usepackage{graphicx}
\usepackage{txfonts}
\usepackage{natbib}
\usepackage{longtable,lscape}
\usepackage{amssymb}
\usepackage{mathrsfs}
\usepackage{amsfonts}
\usepackage{url}
\usepackage{color}

\usepackage[flushleft]{threeparttable}

\def\Planck{\textit{Planck}}
\def\deg{\ifmmode^\circ\else$^\circ$\fi}
\def\pdeg{\ifmmode $\setbox0=\hbox{$^{\circ}$}\rlap{\hskip.11\wd0 .}$^{\circ}
          \else \setbox0=\hbox{$^{\circ}$}\rlap{\hskip.11\wd0 .}$^{\circ}$\fi}
\def\arcs{\ifmmode {^{\scriptstyle\prime\prime}}
          \else $^{\scriptstyle\prime\prime}$\fi}
\def\arcm{\ifmmode {^{\scriptstyle\prime}}
          \else $^{\scriptstyle\prime}$\fi}
\newdimen\sa  \newdimen\sb
\def\parcs{\sa=.07em \sb=.03em
     \ifmmode \hbox{\rlap{.}}^{\scriptstyle\prime\kern -\sb\prime}\hbox{\kern -\sa}
     \else \rlap{.}$^{\scriptstyle\prime\kern -\sb\prime}$\kern -\sa\fi}
\def\parcm{\sa=.08em \sb=.03em
     \ifmmode \hbox{\rlap{.}\kern\sa}^{\scriptstyle\prime}\hbox{\kern-\sb}
     \else \rlap{.}\kern\sa$^{\scriptstyle\prime}$\kern-\sb\fi}
\def\mo{\ifmmode^{-1}\else$^{-1}$\fi}

\begin{document}

\title{Optical validation and characterization of \Planck\ PSZ1 sources at the 
Canary Islands observatories. I. First year of ITP13 observations}
\titlerunning{Optical validation of \Planck\ PSZ1 unknown sources}

\author{R.~Barrena \inst{1,2} \and A.~Streblyanska \inst{1,2} \and A.~Ferragamo \inst{1,2} \and
J.A.~Rubi\~{n}o-Mart\'{\i}n \inst{1,2} \and A.~Aguado-Barahona \inst{1,2} \and D.~Tramonte \inst{1,2,3}
\and R.T.~G\'enova-Santos \inst{1,2} \and A.~Hempel \inst{4} \and H. Lietzen \inst{5} \and N. Aghanim 
\inst{6} \and M. Arnaud \inst{7,8} \and H. B\"ohringer \inst{8} \and G. Chon \inst{8} \and J. Democles 
\inst{7,8} \and H. Dahle \inst{10} \and M. Douspis \inst{6} \and A.~N. Lasenby \inst{11,12} \and P. Mazzotta 
\inst{13} \and J.B.~Melin \inst{7,8} \and E. Pointecouteau \inst{14,15} \and G.W. Pratt \inst{7,8}
\and M. Rossetti \inst{16} \and R.F.J.~van der Burg \inst{7,8,17}}

\institute{Instituto de Astrof\'{\i}sica de Canarias, C/V\'{\i}a L\'{a}ctea s/n, E-38205 La Laguna, Tenerife, Spain\\
\email{rbarrena@iac.es} 
\and
Universidad de La Laguna, Departamento de Astrof\'{i}sica, E-38206 La Laguna, Tenerife, Spain
\and
University of KwaZulu-Natal, Westville Campus, Private Bag X54001, Durban 4000, South Africa
\and
Universidad Andr\'es Bello, Departemento de Ciencias F\'{i}sicas, 7591538 Santiago de Chile, Chile
\and
Tartu Observatory, University of Tartu, 61602 T\~oravere, Tartumaa, Estonia
\and 
Institut d'Astrophysique Spatiale, Universit\`e Paris-Sud, CNRS, UMR8617, 91405 Orsay Cedex, France 
\and
IRFU, CEA, Universit\'e Paris-Saclay, F-91191 Gif-sur-Yvette, France
\and
Universit\'e Paris Diderot, AIM, Sorbonne Paris Cit\'e, CEA, CNRS, F-91191 Gif-sur-Yvette, France
\and
Max-Planck-Institut f\"ur extraterrestrische Physik, D-85748 Garching, Germany
\and
Institute of Theoretical Astrophysics, University of Oslo, PO Box 1029, Blindern, 0315 Oslo, Norway
\and
Astrophysics Group, Cavendish Laboratory, JJ Thomson Av., Cambridge, CB-3 $\Theta$HE, UK
\and
Kavli Institute for Cosmology, Madingley Road, Cambridge, CB3 $\Theta$HA, UK 
\and 
Dipartimento di Fisica, Universit\`a degli Studi di Roma ``Tor Vergata'', via della Ricerca Scientifica, 1, I-00133 Roma, Italy
\and
Universit\'e de Toulouse, UPS-OMP, Institut de Recherche en Astrophysique et Plan\'etologie (IRAP), 31400 Toulouse, France
\and
CNRS, IRAP, 9 avenue Colonel Roche, BP 44346, 31028 Toulouse Cedex 4, France
\and
IASF-Milano, Istituto Nazionale di Astrofisica, via Bassini 15, 20133 Milano, Italy
\and
European Southern Observatory, Karl-Schwarzschild-Str. 2, 85748 Garching, Germany}

\date{Received ; accepted } 

\authorrunning{Barrena et al.}


\abstract{We identify new clusters and characterize previously unknown \Planck\ 
Sunyaev--Zeldovich (SZ) sources from the first \Planck\ catalogue of SZ sources 
(PSZ1). The results presented here correspond to an optical follow-up observational programme 
developed during approximately one year (2014) at Roque de los Muchachos Observatory, using the 
2.5 m Isaac Newton telescope, the 3.5 m Telescopio Nazionale Galileo, the 4.2 m William Herschel 
telescope and the 10.4 m Gran Telescopio Canarias. We characterize 115 new PSZ1 sources using 
deep optical imaging and spectroscopy. We adopt robust criteria in order to consolidate the 
SZ counterparts by analysing the optical richness, the 2D galaxy distribution, and velocity 
dispersions of clusters. Confirmed counterparts are considered to be validated if they are rich 
structures, well aligned with the \Planck\ PSZ1 coordinate and show relatively high 
velocity dispersion. Following this classification, we confirm 53 clusters, which means 
that 46\% of this PSZ1 subsample has been validated and characterized with this 
technique. Sixty-two SZ sources (54\% of this PSZ1 subset) remain unconfirmed. In addition, we 
find that the fraction of unconfirmed clusters close to the galactic plane (at $|b|<25\deg$) 
is greater than that at higher galactic latitudes ($|b|>25\deg$), which indicates contamination
produced by radio emission of galactic dust and gas clouds on these SZ detections. In fact, in 
the majority of the cases, we detect important galactic cirrus in the optical images, mainly 
in the SZ target located at low galactic latitudes, which supports this hypothesis.}


\maketitle

\section{Introduction}

The Sunyaev--Zeldovich (SZ) effect \citep{sz1972} is a spectral distortion
of the cosmic microwave background (CMB) generated by high energy 
electrons interacting with hot CMB photons through inverse Compton scattering. The SZ effect
has in recent years become a powerful tool in cosmology that can complement 
the information obtained from the CMB angular power spectrum 
\citep[e.g.][]{birkinshaw1999, carlstrom2002, planck2013-p15, planck2013-p05b,
planck2013-p05a}. Nowadays, one of the most common applications of the SZ effect 
is the detection of galaxy clusters. These gravitationally bound systems emerge as 
massive structures in the cosmic web of the large-scale structure 
\citep[e.g.][]{springel05}. Galaxy clusters encompass several components, including 
dark and baryonic matter \citep[e.g.][]{allen11}. For this reason, galaxy 
clusters are excellent laboratories for testing cosmology and establishing constraints on 
cosmological parameters such as dark matter, dark energy densities, and the equation 
of state of the dark energy and neutrino masses \citep{vikhlinin09, Henry2009, 
mantz10a, planck2013-p15, mantz2015, planck2014-a30}. 

The SZ effect is particularly evident when CMB photons interact, following the 
inverse Compton mechanism, with massive haloes that have a high content of hot gas. The net
effect on the initial Planck spectrum of the CMB is to shift it to higher frequencies.
While the surface brightness of the SZ effect is independent of redshift,  
cluster counts at high redshift are very sensitive to the cosmology (see e.g. 
\citealt{borgani01}). Nevertheless, these massive structures are predicted to be very 
scarce in the $\Lambda$CDM model, and their abundance strongly decreases at high redshift 
\citep[e.g.][]{springel05}. Therefore, the detection of galaxy clusters via the SZ effect, 
now accessible also with all-sky surveys, can be used for surveying large volumes and for 
constraining cosmological parameters.

In recent decades, all-sky surveys have been conducted using either optical or X-ray
observations. Examples of optical surveys include catalogues based on SDSS data
\citep{koester2007b,wen2009,hao2010,szabo2010,WHL2012}, which are complete up to $z\sim 0.3$
for galaxies with magnitudes $M_r<-22$, reaching a limit of about $z\sim 0.5$. Examples of X-ray 
surveys are the ROSAT All-Sky Survey (RASS) and 
their corresponding catalogues (e.g.\ REFLEX \citep{reflex,bohringer2014}, MACS \citep{macs}, and 
NORAS \citep{noras}), which extend to similar redshifts. Concerning millimetre 
surveys, in the last few years, remarkable efforts have yielded the first complete 
lists of galaxy clusters compiled through the thermal SZ effect, such as the surveys 
carried out with the Atacama Cosmology Telescope (ACT; \citealt{Marriage2011}; 
\citealt{Hasselfield2013}) and the South Pole Telescope (SPT; 
\citealt{Staniszewski2009}; \citealt{Vanderlinde2010}; \citealt{Williamson2011}; 
\citealt{Reichardt2013}; \citealt{Bleem2015}).

The \Planck\footnote{\Planck\ (\url{http://www.esa.int/Planck}) is a project of
  the European Space Agency (ESA) with instruments provided by two scientific
  consortia funded by ESA member states and led by Principal Investigators from
  France and Italy, telescope reflectors provided through a collaboration
  between ESA and a scientific consortium led and funded by Denmark, and
  additional contributions from NASA (USA). } satellite \citep{planck2013-p01}
provided the first opportunity to detect galaxy clusters through
the SZ effect in a full sky survey
\citep{planck2011-5.1a,planck2013-p05a,planck2014-a36}. Nevertheless, some SZ 
detections may correspond to misidentifications (in particular those with low
SZ signal) or contamination due to the galactic dust contribution. In addition, 
the SZ effect provides no information about the redshift of the clusters. So, for 
all these reasons, dedicated follow-up programmes are important to make the 
resulting catalogues scientifically useful. In 2010, the \Planck\ collaboration 
started intensive follow-up programmes to confirm SZ cluster candidates, first
from intermediate versions of the \Planck\ SZ catalogue 
\citep{planck2011-5.1b,planck2012-I,planck2012-IV}, second from the first 
public SZ catalogue (PSZ1; \citealt{planck2013-p05a,planck2013-p05a-addendum}), and
finally from the second public SZ catalogue (PSZ2; \citealt{planck2014-a36}). Examples 
of optical follow-up of unknown SZ candidates are the observational programme
performed with the RTT150 telescope \citep{planck2014-XXVI}, that carried out at the
Canary Islands observatories \citep{planck2016-XXXVI}, and the validation based
on MegaCam at the 3.6 m Canadian France Hawaii Telescope \citep{megacam} (hereafter vdB+16). 
The RTT150 programme confirmed 79 PSZ1 clusters (47 of which were previously unknown), thereby
determining photometric and spectroscopic redshift in the range 
$0.08<z<0.83$. This technique is also followed in \citet{planck2016-XXXVI}, confirming
73 PSZ1 clusters and providing their redshifts (in the range $0.09<z<0.82$). They also
report no counterparts found for five additional cases. Finally, \citet{megacam} 
investigated a sample of 28 PSZ1 and PSZ2 candidates that were pre-selected using 
WISE and SDSS as high redshift ($z>0.5$) candidates, with deep $r$- and $z$-band imaging 
from MegaCam/CFHT.

In this paper, we continue the optical characterization of PSZ1 SZ sources started in 
\citet{planck2016-XXXVI}. The main motivation of these studies is to validate and 
identify cluster counterparts of unknown PSZ1 sources.

The present paper is organized as follows. In Sect.~\ref{sec:sample} we  describe 
the \Planck\ PSZ1 catalogue and detail the observational follow-up. In Sect.~\ref{sec:photoz} 
we describe the technique used to identify and confirm the cluster counterparts. 
Sect.~\ref{sec:results} includes a detailed description of the nature of some relevant SZ 
targets (multiple detections, presence of gravitational arcs, fossil systems, etc.). Finally, 
in Sect.~\ref{sec:conclusions} we present the conclusions.


\section{The PSZ1 and optical follow-up observations}
\label{sec:sample}

Our reference cluster sample is the first \Planck\ catalogue of SZ sources (PSZ1;
\citealt{planck2013-p05a,planck2013-p05a-addendum}). This catalogue includes 1227
clusters and cluster candidates derived from SZ effect detections using all-sky
maps produced within the first 15.5 months of \Planck\ observations. Briefly,
SZ sources are selected using three different detection methods: MMF1, MMF3, 
and PwS. All sources included in the PSZ1 catalogue were detected by at least one of these three detection methods with a signal-to-noise 
ratio (S/N) of 4.5 or higher. 
\citet{planck2013-p05a} describes in detail the selection method followed in the 
construction of the catalogue. The PSZ1 contains redshifts for 913 systems, of 
which 736 are spectroscopic. The purity depends on the S/N of the SZ detection, and is 
estimated to be $\sim$95\% for the entire PSZ1 sample.

The second \Planck\ SZ catalogue (PSZ2) is already publicly available \citep{planck2014-a36}.
However, the optical follow-up presented here includes only PSZ1 sources 
(which will be targets for a forthcoming new study). Our long-term observing programme
has recently been finished and we are analysing the data retrieved.

An initial effort at validating the PSZ1 sources \citep{planck2013-p05a} was made 
in order to search for SZ counterparts in optical data through the 
DSS.\footnote{DSS: \url{http://stdatu.stsci.edu/dss}} images, SDSS survey 
\citep[SDSS, DR8,][]{2011ApJS..193...29A} and the {\it WISE} all-sky survey \citep{wise}
In addition, the PSZ1 catalogue was cross-correlated with X-ray data, mainly with 
the ROSAT All Sky Survey \citep[RASS,][]{rassbr,rassfaint}, as well as other SZ catalogues.
After these processes, the unconfirmed SZ sources were targeted for follow-up 
observations at the RTT150 
telescope \citep{planck2014-XXVI}, the Canary Islands observatories (this work and 
\citealt{planck2016-XXXVI}), and the MegaCam at CFHT \citep{megacam}. 
\citet{alina2018} have very recently characterized 37 new \Planck\ PSZ2 targets and presented updated redshifts
for a sample of PSZ1 targets using the SDSS DR12 database.

\subsection{Optical follow-up observations}
\label{sec:followup}

All the observations were carried out at Roque de los Muchachos Observatory
(ORM) on the island of La Palma (Spain) within the framework of the International Time Programme ITP13B-15A. 
The dataset was obtained in multiple runs from August 2013 to July 2014, as part of 
this two-year observing programme.

Table~\ref{tab:telescopes} lists the ORM instruments and telescopes used for this 
follow-up work: a) the 2.5 m Isaac Newton Telescope and 
the 4.2 m William Herschel Telescope operated by the Isaac Newton Group of Telescopes (ING);
b) the 3.5 m Italian Telescopio Nazionale Galileo (TNG) operated by the Fundaci\'on 
Galileo Galilei of the INAF (Istituto Nazionale di Astrofisica), and c) the 10.4 m Gran 
Telescopio Canarias (GTC) operated by the Instituto de Astrof\'{\i}sica de Canarias 
(IAC).

Our targets are unknown PSZ1 clusters, which usually correspond to low S/N SZ  
sources in this catalogue. However, the observational strategy prioritizes targets with 
highest SZ S/N, with the sole restriction that targets have declinations $>-15\deg$, in 
order to be observable from the ORM. Targets with lower declinations are scheduled for
other follow-up programmes to be developed from Southern Hemisphere facilities.

\begin{table*}[h!]
\begingroup
\newdimen\tblskip \tblskip=5pt
\caption{Telescopes and instruments at ORM used in this optical follow-up 
validation programme. Columns 3 to 5 show field of view, the pixel scale and the resolution 
used in the imaging and spectroscopic observing mode. The last two columns list the total number 
of clusters observed performing imaging and spectroscopy.}
\label{tab:telescopes}
\nointerlineskip 
\vskip -2mm 
\footnotesize
\newdimen\digitwidth
\setbox0=\hbox{\rm 0}
\digitwidth=\wd0
\catcode`*=\active
\def*{\kern\digitwidth}
\newdimen\signwidth
\setbox0=\hbox{+}
\signwidth=\wd0
\catcode`!=\active
\def!{\kern\signwidth}
%
\halign{\hbox to 1.3in{#}\tabskip=2em&
        \hfil#\hfil& 
        \hfil#\hfil&
        \hfil#\hfil&
        \hfil#\hfil&
        \hfil#\hfil&
        \hfil#\hfil&
        \hfil#\hfil\tabskip=0pt\cr
\omit\hfil Telescope & Instrument & FoV & Pixel Scale
           [$\arcs$]& Resolution & N$_{\textrm{ima}}$ & N$_{\textrm{spec}}$ \cr
\noalign{\vskip 3pt\hrule\vskip 5pt}
2.5m  INT&      WFC  & $34\arcm \times 34\arcm$    & 0.33* &   --    & 42 & -- \cr
3.6m  TNG&  DOLORES  & $8\parcm6 \times 8\parcm6$  & 0.252 & $R=600$ & *1 & 20 \cr
4.2m  WHT&     ACAM  & $8\arcm \times 8\arcm$      & 0.253 & $R=400$ & 71 & *2 \cr
10.4m GTC&   OSIRIS  & $7\parcm8 \times 7\parcm 8$ & 0.254 & $R=500$ & -- & 26 \cr
\noalign{\vskip 5pt\hrule\vskip 3pt}}
\endgroup
\end{table*}

We searched for possible counterparts in the Sloan Digital Sky Survey (SDSS)\footnote{\url{http://skyserver.sdss.org}} 
and the Digitized Sky Survey (DSS).\footnote{\url{http://archive.stsci.edu/dss}} After
this previous screening, and depending on whether the cluster was confirmed as counterpart, new imaging was
not needed and only spectroscopic observations were required. Galaxy cluster members with
SDSS spectroscopic information were also considered in order to compute the mean cluster
redshift. 

In short, after previous screening in archive and public data searches for possible
SZ counterparts, our observational strategy followed two steps. First, cluster 
counterparts were identified using deep images and, if these existed (as galaxy overdensities), 
they were studied photometrically using $g^\prime$, $r^\prime$, and $i^\prime$ broad band 
filters. Second, clusters were definitely confirmed through spectroscopic observations, 
either using long-slit or multi-object spectroscopy (MOS). Finally, taking into account all the
photometric and spectroscopic information, the cluster validation was then performed based 
on the selection criteria detailed in Section~\ref{sec:photoz}.

\subsubsection{Imaging observations and data reduction}

Imaging observations were carried out using the Wide-Field Camera (WFC)
mounted on the 2.5\,m INT and the auxiliary-port camera (ACAM) mounted on the 4.2\,m
WHT. We obtained images for every target in the $g^\prime$, $r^\prime$ and $i^\prime$ Sloan bands. 
The WFC camera at the INT is a four EEV $2\,{\rm k}\times 4\,{\rm k}$ CCD mosaic with
a projected pixel scale of $0\parcs33$, giving a field of view (FOV) of $34\arcm 
\times 34\arcm$. ACAM is mounted at the folded Cassegrain focus of the 4.2\,m WHT, 
which provides a FOV of $4\arcm$ radius with a projected pixel scale of $0\parcs25$.

The High Frequency Instrument (HFI) \Planck\ maps extend from 100 to 857\,GHz,
and their beam FWHM varies from $9\parcm6$ at the lowest frequencies to $4\parcm5$ at
the highest. The positional error of sources is about 
$2\arcm$ for targets in the PSZ1 sample \citep{planck2013-p05a}. This result 
has been confirmed by comparing \Planck\ SZ and REFLEX II sources. 
For 83 clusters that overlap in the two catalogues, \citet{bohringer2013} found that 78\% 
are found with a detection offset smaller than $2\arcm$. Optical 
counterparts are therefore not expected to be found beyond 2.5 times the beam size, which means 
that the cluster associated with the SZ effect should be closer than $\sim 5\arcm$ from 
the corresponding PSZ1 coordinate. However, when the clusters are nearby systems 
($z<0.2$) their apparent radius may fill a large region, and in these cases their centre 
offsets relative to their SZ position may be higher. So a typical field of $10\arcm$ 
is large enough to cover the region where counterparts are expected with respect to the 
nominal SZ \Planck\ coordinates \citep{planck2013-p05a}.

The fields observed using the WFC were acquired with 1500\,s exposures per band and by performing 
a small dithering pattern of three points with about $10\arcs$ offset. This technique 
allowed us to remove bad pixels and vignetting, correct most of the fringing effects that 
are present in the CCD, and minimize the effect of cosmic rays. An analogous
procedure was applied in the ACAM acquisitions, with the only difference that the total 
integration time per band was 900\,s, split into three separate exposures of 
300\,s. The completeness and limit magnitudes\footnote{Completeness and limit magnitudes 
correspond to detection targets with $S/N\sim5$ and 3, respectively.} 
computed from object counts in the $r^\prime$-band obtained with these integration times 
were $r^\prime=23.2$ and 24.0 mag using the WFC, and $r^\prime=23.8$ and 24.9 mag in the ACAM 
frames, respectively.

The optical WFC and ACAM images were reduced using standard {\tt IRAF} tasks.\footnote{{\tt IRAF} 
(\url{http://iraf.noao.edu/}) is distributed by the
  National Optical Astronomy Observatories, which are operated by the
  Association of Universities for Research in Astronomy, Inc., under cooperative
  agreement with the National Science Foundation.} The reduction
procedure includes bias and flat-field corrections and astrometric calibrations. 
The astrometry was performed using the {\tt images.imcoords} task and the USNO B1.0 catalogue
as reference. The photometric calibration refers to SDSS photometry and SDSS standard 
fields. When the observations were carried out in non-photometric conditions, the targets
were calibrated in the subsequent observing run under clear skies.

The source detection and photometry were performed using {\tt SExtractor} \citep{bertin1996}
in single-image mode. Sources were detected in each $g^\prime$-, $r^\prime$-, 
and $i^\prime$-band if they presented more than 10 pixels with $S/N>1.5 \sigma$ 
detection threshold in the filtered images. We performed elliptical aperture photometry 
with variable size using the {\tt MAGAUTO} set-up. The parameters of the `Kron factor' and 
the `minimum radius' were set to the default values of 2.5 and 3.5, respectively. All the 
photometric catalogues were then merged in order to create a master catalogue containing 
the three-band photometries.

Unfortunately, the WFC presents large PSF distortions over the wide FOV. This implies 
that the star/galaxy classification was very difficult for faint sources with $r^\prime>18$ mag. 
So, we perform star-cleaning procedures within small regions containing the cluster 
candidates in order obtain clear RS in the colour--magnitude diagrams.

\subsubsection{Spectroscopic observations and data reduction}
\label{sec:spec_data_subsec}

The majority of the data presented here were acquired using the DOLORES 
and OSIRIS spectrographs at the 3.5\,m TNG and the 10.4\,m GTC, respectively. However, in 
a few cases, ACAM (at the 4.2\,m WHT) was also used in its spectroscopy mode. DOLORES 
and OSIRIS observing blocks were made in MOS mode while ACAM was set-up in long slit 
configuration.


We used the DOLORES spectrograph at the TNG. This instrument has a $2k\times 2k$ CCD
detector with a pixel scale of $0\parcs252$. We performed low resolution spectroscopy
using the LR-B grism (dispersion of 2.75\,$\AA$\,pixel\mo\ in the range 4000--8000\,$\AA$) with 
slit widths of $1\parcs6$, yielding a resolution of $R\sim 600$. We placed about
40--45 slitlets per mask and acquired He-Ne and Hg-Ne arcs, which allowed us to obtain 
wavelength calibrations with $rms \sim$0.1\,$\AA$ over the whole wavelength range.
We obtained $3\times1800$\,s exposures per cluster in order to obtain spectra
with $S/N \sim 5$ for galaxies with magnitudes $r^\prime=20.5$. 


We used the OSIRIS spectrograph at the GTC in MOS observing mode. This instrument has a double
$2k\times 4k$ CCD detector used with a binning of $2\times 2$ pixels, which gives
a final pixel scale of $0\parcs25$. We placed up to 60 slitlets per mask of $1\parcs2$ 
widths. The R300B grism was used in order to get a dispersion of 5.2\,$\AA$\,pixel\mo\ 
($R \sim 500$) over the full visible wavelength range. We obtained wavelength calibrations
with $rms \sim$0.2\,$\AA$ accuracy using Hg, Ne, and Ar lamps. $3\times1000$\,s exposures
per mask were acquired in order to get spectra with $S/N \sim 5$ for galaxies with
magnitudes $r^\prime=21.6$. For some specific targets we used the OSIRIS spectrograph in 
long-slit mode, using the same grism and slit width set-up as used for the MOS unit.


Both the DOLORES/TNG and OSIRIS/GTC masks used 3--5 pinholes as pivots placed on a fiducial
star to centre the masks. The masks were designed using previous imaging for each field,
and galaxies were selected from the photometry obtained with the WFC. Basically, 
we used RGB colour composite images as a reference and considered cluster members that 
galaxies contained within the RS of the cluster, obtained from $g^\prime-r^\prime$ 
and $r^\prime-i^\prime$ colours. This selection procedure yields a success rate of 
about 60\% in the core of the clusters, and 20\% in the external regions ($>0.3$ Mpc 
from the bright cluster galaxy, the BCG).


We used ACAM at the 4.2\,m WHT telescope. This instrument also offers a long-slit
spectroscopy mode. It is equipped with a $2k\times 2k$ CCD with a projected pixel scale 
of $0\parcs25$. The disperser is a holographic device offering a wavelength coverage 
between 4000 and 9000\,$\AA$, a dispersion of 3.3\,$\AA$\,pixel\mo and a resolution of 
$R\sim 400$ for a slit width of $1\arcs$. We obtained spectra for the BCG and 
several other cluster members by positioning the slit in two or more orientations. We used exposure times of
$2\times 1000$\,s per slit in order to obtain spectra with $S/N\sim 5$ for
galaxies with magnitudes $r^\prime \sim 19.5$. We noticed that this spectrograph is not
so efficient as other homologous devices, probably owing to its holographic disperser. 
We obtained lower $S/N$ spectra as expected for the chosen exposure times.

The data reduction included sky subtraction, extraction of spectra, cosmic ray 
rejection, and wevelength calibration. All these procedures were performed using
standard {\tt IRAF} tasks. After a careful test, we decided to do not apply bias 
and flat-field correction, because these corrections added unwanted noise 
to the low signal of the faint sources. The wavelength calibration of scientific 
spectra was performed using He-Ne, Hg, and Ar arcs. We then checked the calibrated 
spectra and looked for possible deviations using the OI telluric line ($5577.3$\,$\AA$). 
We found no systematic offsets, but random deviations of about $1$\,$\AA$, which 
correspond to $\sim$50\,km\,s$^{-1}$ with the instrument set-up used.

\begin{figure*}[h!]
\centering
\includegraphics[width=\columnwidth]{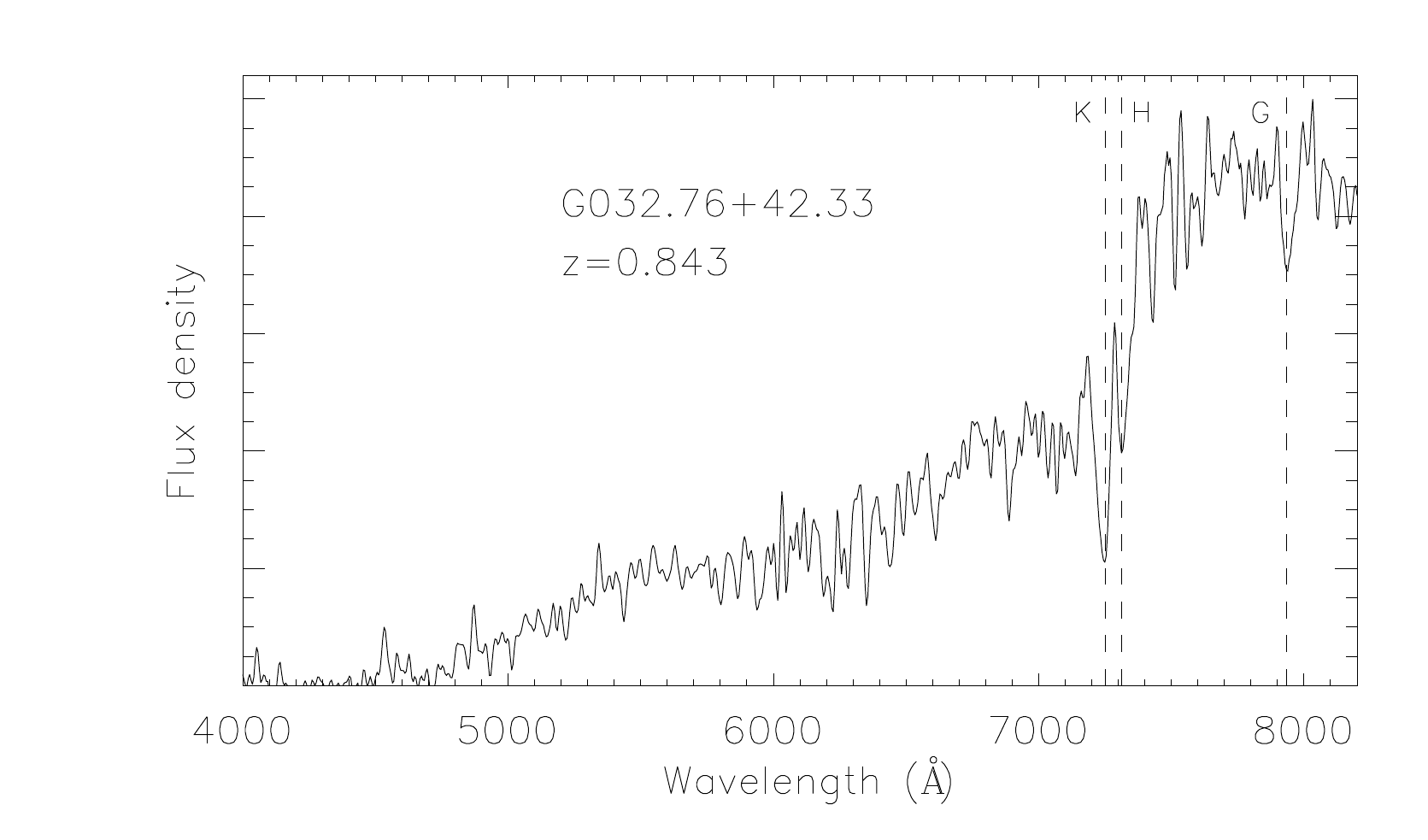}
\includegraphics[width=\columnwidth]{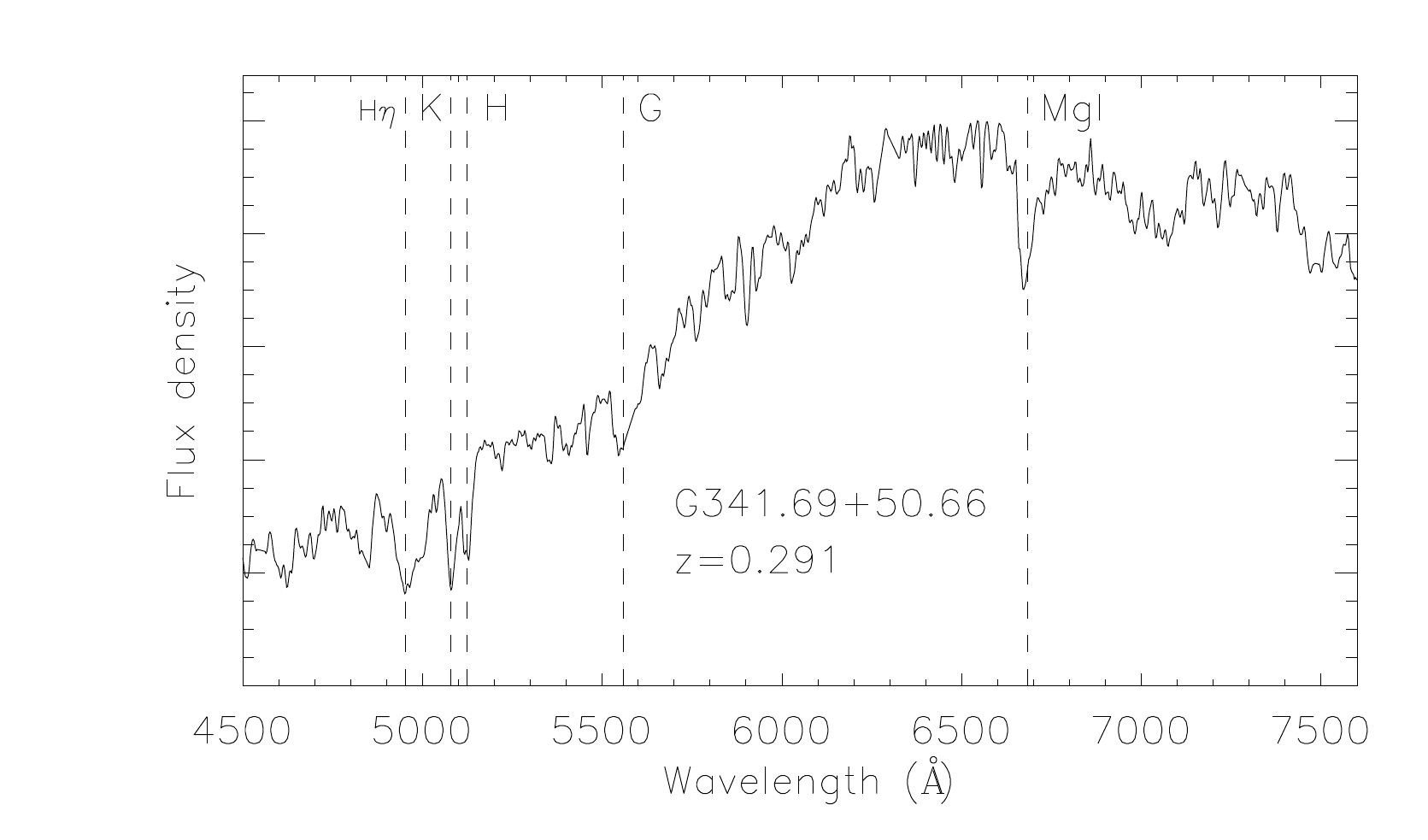}
\caption{\color{black} Example of the spectra obtained with GTC/OSIRIS (left panel) 
and TNG/DOLORES (right panel) for two luminous galaxy members, with magnitudes
$r^\prime$ = 21.6 and 17.8, in the clusters G032.76$+$42.33 and 
G341.69$+$50.66, at $z=0.843$ and 0.291 respectively. Dashed lines correspond to the wavelength of the absorption features identified in 
each spectrum at the redshift of the clusters. Flux density is plotted in arbitrary units.}
\label{fig:xc_spec}
\end{figure*}

The reduced spectra showed $S/N\sim5$ (per pixel around $5500$\,$\AA$)
for galaxies with $r^\prime \sim 19.5, 20.5$, and $21.7$ mag, observed with
the WHT, TNG, and GTC respectively. Our observational strategy consisted in observing the 
nearby clusters ($z_{\rm phot} \lesssim 0.4$) at the TNG and WHT, whereas the most 
distant systems ($z_{\rm phot} \gtrsim 0.4$) were observed using the GTC. 
Figure~\ref{fig:xc_spec} shows two examples of spectra obtained at the TNG and GTC.

The long-slit observations were planned in order to get the maximum number of redshift
estimates by placing between two and four galaxies within the slit, and selecting two 
position angles for each cluster candidate. We always included the BCG in one
of the position angles. Therefore, with this scheme we were able to obtain as many as
5--6 redshifts per cluster candidate.

We estimated radial velocities using the cross-correlation technique
\citep{tonrydavis79} implemented in the {\tt IRAF} task {\tt
RVSAO}.\footnote{RVSAO was developed at the Smithsonian Astrophysical
Observatory Telescope Data Center.} The science spectra were 
correlated with six spectrum templates \citep{kennicutt92} of various galaxy
morphologies (E, S0, Sa, Sb, Sc, and Irr). The {\tt XCSAO} procedure provides an
$R$ parameter linked with the $S/N$ ratio of the cross-correlation peak. We chose
the radial velocity corresponding to the template with the highest $R$-value. 
Most of the redshifts were estimated using only absorption lines (mainly the H and 
K CaI doublet, G-band, and MgI triplet), as corresponds to a dominant early-type 
population of galaxies in clusters. However, a few cases showed strong
emission lines (mainly OII, OIII doublet, and H$_\beta$), which were used to 
determine the redshift. Finally, we visually inspected all spectra to verify
the velocity determination, which was particularly difficult for galaxies at 
$z>0.7$, owing to the low $S/N$ of the continuum of the spectra and the few 
spectral features identified within the visible wavelength range. 

At the end of this process, the cross-correlation yielded a mean error in the radial 
velocity estimates of $\Delta$v$\sim 75$\,km\,s$^{-1}$. However, double redshift 
determinations for a set of about 50 spectra allowed us to estimate the true 
intrinsic errors. By comparing the first and second velocity estimates we obtained
an rms of $\Delta v=110$\,km\,s$^{-1}$. This means that actual errors in the
redshift estimates are $\Delta z \sim 0.0004$, which is in agreement with the 
predicted value for the resolution set-ups of the spectrographs used in this study.

The multi-object masks allowed us to sample the core of the clusters in a more 
complete way. We obtained typically 40--50 radial velocities per mask and retrieved 10--25 
galaxy members per cluster, which means a success rate of about 20--50\%. We 
estimated the mean redshift and velocity dispersion of the clusters. In all cases,
the cluster redshift was assumed to be the mean value for the galaxy members retrieved.

In this study, we considered galaxy members only if they showed radial velocities of 
$\pm 2500$\,km\,s$^{-1}$ with respect to the mean velocity of the systems. Given that this
range is about three times the typical velocity dispersion of a cluster, this
criterion guaranteed that we selected the vast majority of galaxy members while
minimizing contamination of interlopers.


\section{Cluster identification and confirmation criteria}
\label{sec:photoz}

\begin{figure}[ht!]
\centering
\includegraphics[width=\columnwidth]{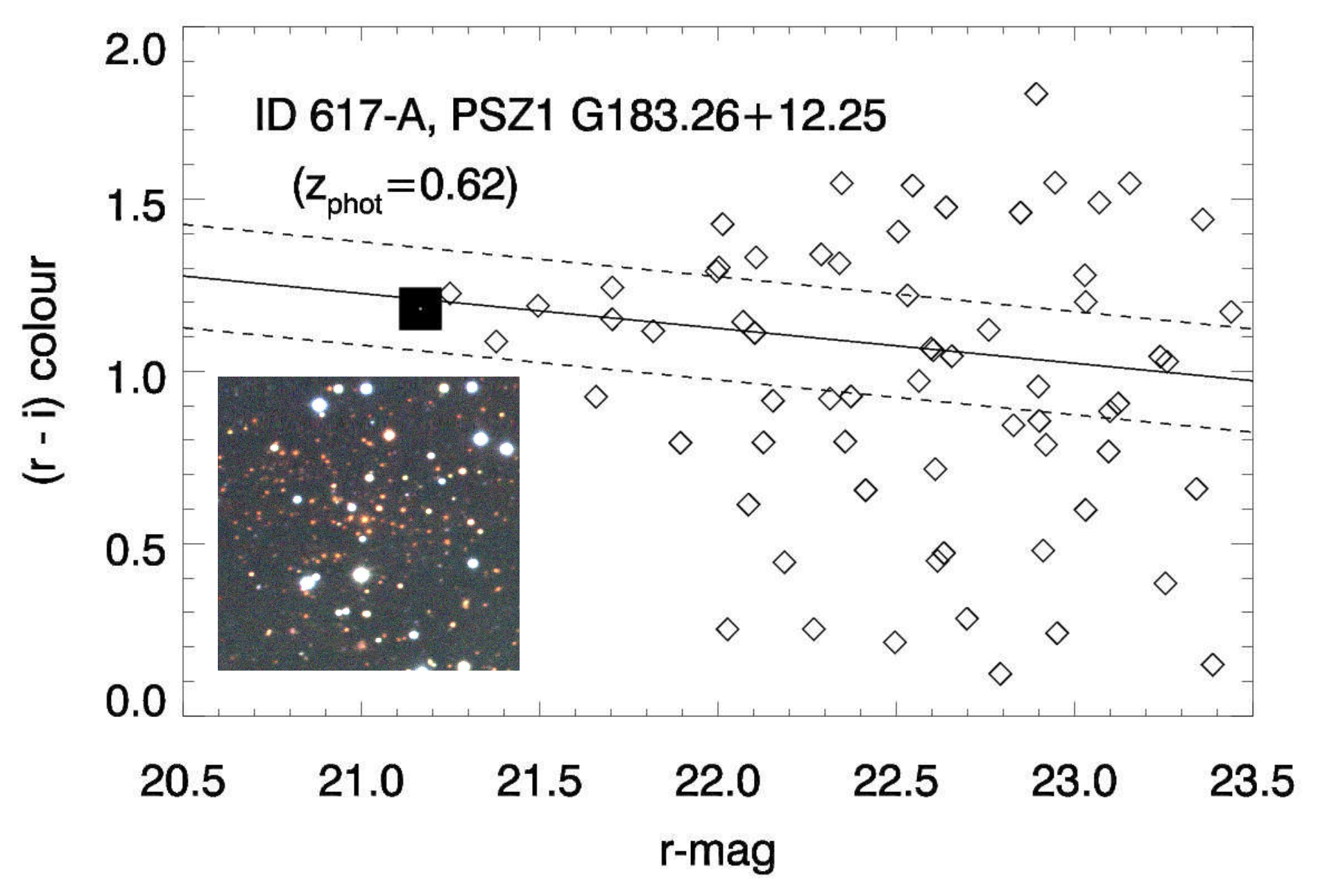}
\caption{($r^\prime - i^\prime$,$r^\prime$) CMD of ID 617-A, associated with PSZ1 the source 
G183.26$+$12.25. The diagram considers extended sources within a region of $2\parcm5 \times 2\parcm5$ 
($1 \times 1$ Mpc) around the BCG of the system (represented by the large filled square in the 
plot). The solid line represents the fitted RS, which yields  $z_{\rm phot}=0.62$. Dashed lines 
correspond to the RS$\pm$0.15 locus, where likely cluster members are selected. An RGB 
image (composed using the $g^\prime$, $r^\prime$ and $i^\prime$ frames) showing the $2\parcm5 
\times 2\parcm5$ inner region of this cluster is superimposed on the plot.}
\label{fig:cmd}
\end{figure}

We identified clusters in the images as galaxy overdensities showing coherent colours. 
The significance of each overdensity is evaluated through the richness parameter $R$ (as
defined in Sect.~\ref{sec:criteria}), whereas 
colours and photometric redshifts are estimated on the basis of the red sequence (hereafter RS; 
\citealt{gladders2000}). We selected likely galaxy members using the cluster 
RS in the $(g^\prime - r^\prime, r^\prime)$ and $(r^\prime -i^\prime, r^\prime)$ 
colour--magnitude diagrams (CMD). We define the RS as galaxies showing a coherent 
colour. That is, we look for galaxies with similar colours around a bright one (assumed to be 
the BCG) in the observed region forming a compact group of galaxies in space. We fit 
the RS by calculating the mean colour of the five brightest likely members and fixing the RS 
slope to $-0.039$ and $-0.017$ in $(g^\prime - r^\prime, r^\prime)$ and $(r^\prime -i^\prime, 
r^\prime)$ CMDs respectively (see \citealt{Barrena2012}). Thus, we combined the colour 
information of galaxy member candidates and their spatial distribution in order to search for 
galaxy overdensities, thus estimating the corresponding photometric redshift (see 
Fig.~\ref{fig:cmd}). We use the mean colour of the five brightest galaxies of the 
cluster to estimate the photometric redshift. The process followed to estimate 
$z_{\rm phot}$ using the colours of the galaxy population is detailed in 
\citeauthor{planck2016-XXXVI} (see section 4.2 and eqns 1 and 2 therein). In 
addition, galaxies with colours within the RS$\pm$0.15 were considered as cluster member 
candidates and selected as galaxy targets for further spectroscopic observations.

We also use the $g^\prime$-, $r^\prime$-, and $i^\prime$-band images in order to compose RGB colour 
images. These RGB (colour composite) frames were used to consolidate our findings. Therefore, 
actual galaxy overdensities were clearly identified by eye. This visual inspection proved 
to be very efficient for the identification of high redshift ($z>0.7$) clusters, or even fossil systems. 
In the former, our optical data were not deep enough to identify the cluster RS, but the galaxy 
overdensities corresponding to the brightest members were clearly visible in the RGB images at 
the minimum detection level. On the other hand, fossil systems \citep[e.g.][]{jones2003,voevodkin2010}
are very evolved structures with a mass content capable of producing a detectable SZ effect on \Planck\ 
maps \citep{pratt2016}. These kinds of galaxy clusters, characterized by the presence of a huge central 
BCG and a poor galaxy population, are also particular cases that are not easily detected by automatic 
algorithms in colour space. Their RS are usually not very populated owing to the 2 magnitude 
gap beyond the BCG, so these kinds of clusters do not fit the standard RS. Visual 
inspection of RGB images is therefore essential.

\subsection{Confirmation criteria}
\label{sec:criteria}

In order to perform a robust confirmation of the PSZ1 sources, we adopt the criteria set out 
in Table \ref{tab:criteria}. This set of conditions is based on the dynamical properties of 
the clusters (through their velocity dispersion retrieved from spectroscopic observations), 
richness, and distance from the \Planck\ pointing. We therefore classify the counterparts 
with different flags according to the validation level of each target.

\begin{table}[h!]
\caption{Confirmation criteria adopted to validate or reject clusters as counterparts of 
SZ detections.}
\label{tab:criteria}
\begin{center}
\begin{tabular}{|c|c|l|c|c|}
\hline
{\tt Flag} & Spectroscopy & $\sigma_v$ limit  & R	    & Dist.  \\ 
           &	          & (km\,s$^{-1}$)    & (N$_{gal})$ &        \\ 
\hline
1  & YES & $>500 \ ; \ 0<z<0.2$ & $>15$  & $<5\arcm$  \\  
   &     & $>650 \ ; \ z>0.2$   & $>15$  & $<5\arcm$  \\  
\hline
2  & NO  & NA                   & $>15$  & $<5\arcm$  \\  
\hline
3  & YES & $<500 \ ; \ 0<z<0.2$ & $>15$  & $<5\arcm$  \\  
   &     & $<650 \ ; \ z>0.2$   & $>15$  & $<5\arcm$  \\  
   & NO  & $-$                  & $>15$  & $>5\arcm$  \\  
\hline 
ND & $-$ & $-$                  & $\le 15$ & $-$     \\  
\hline
\end{tabular}
\end{center}
\end{table}

The mass--redshift distribution of the \Planck\ cluster sample is reported in 
\citet{planck2013-p05a-addendum}, which provides the detectable mass of SZ sources in the 
PSZ1 catalogue. So, following this relation, one would expect that no poor systems are
validated even if they are in the SZ line of sight. However, owing to the noise inhomogeneity in 
the \Planck\ $Y_{500}$ maps (\citet{planck2013-p15}; see Fig. 4 therein), or statistical effects, 
relatively low-mass haloes may scatter beyond the SZ detection threshold. The statistical influence 
on detection samples is known as the Eddington bias \citep{eddington1913}. vdB+16
studied this effect on the measured SZ-based masses and confirmed 13 SZ counterparts at $z>0.5$
showing M$_{500} \gtrsim 2 \ 10^{14}$M$_\odot \ h_{70}^{-1}$. Moreover, nearby ($z<0.2$) galaxy 
systems may also be detected in \Planck\ SZ maps, some even showing masses M$_{500} \sim 10^{14}$M$_\odot 
\ h_{70}^{-1}$. Given that we have spectroscopic information for a reasonable number of members 
and clusters, we can determine the velocity dispersions, and use them to investigate whether they are
poor or massive systems. According to the M$_{200}$-$\sigma_v$ scaling adopted in \citet{munari2013} 
and the M$_{200}$-M$_{500}$ relation suggested by \citet{komatsu2011} (see eq.\ D15 therein), 
we estimate that clusters at $z<0.2$ with M$_{500} > 10^{14}$M$_\odot \ h_{70}^{-1}$ should 
present $\sigma_v > 500$~km\,s$^{-1}$, whereas clusters at $z>0.2$ with M$_{500} > 2 \times 
10^{14}$M$_\odot \ h_{70}^{-1}$ should show $\sigma_v > 650$~km\,s$^{-1}$. So we assume these 
limits on $\sigma_v$ in order to distinguish between actual and detectable systems by \Planck\ 
and chance identifications not linked to the SZ effect.

$R$ is the `richness' computed as the number of likely members (galaxies in the RS$\pm0.15$\,mag 
locus, which represents the $\pm 3 \sigma$ colour dispersion of RS at its brightest 
part) in $g^\prime$-$r^\prime$ and $r^\prime$-$i^\prime$ for clusters at $z \leq 
0.35$ and $z>0.35$ respectively, showing magnitudes in the range [$r^\prime_{BCG}$, 
$r^\prime_{BCG}+3$] within a projected region of 0.5 Mpc radius from the centre of the 
cluster (assumed to be the BCG) at the redshift of the cluster. Likely member counts
were decontaminated from the field contribution, which was computed considering the 
full set of images, excluding the 0.5 Mpc radius regions containing the core of the clusters.
So the richness was obtained from likely member counts after subtraction of a mean field 
contribution (as a function of the redshift).

Cluster optical centres are taken as the BCG position, and centres of clusters that show no 
clear BCG are supposed to be the position of the most luminous member (or likely cluster member
in cases where we have only photometric information). We also adopt confirmation criteria based on the 
distance to the PSZ1 coordinates. The predicted centre error for the whole PSZ1 is $2\arcm$ 
\citep{planck2013-p05a}, so we consider as valid optical counterparts only those clusters at $<5\arcm$ 
from the nominal position, which is a $>2.5\sigma$ position error with respect to the SZ peak emission.

Therefore, following these three criteria based on velocity dispersion, richness and distance 
(see table \ref{tab:criteria}), definitely confirmed optical counterparts are classified with 
{\tt Flag}$=1$, which are massive, rich, and well centred systems (this subset includes ACO and 
Zwiky clusters); clusters potentially validated are classified as {\tt Flag}$=2$, which includes 
clusters well aligned with the \Planck\ pointing with a few galaxy members spectroscopically 
confirmed or targets with no spectroscopy yet available (NA$\equiv$ `Not available' in table 
\ref{tab:criteria}) but rich from the optical point of view; and {\tt Flag}$=3$ corresponds to 
clusters very marginally associated with their corresponding SZ signal owing to their low $\sigma_v$ 
or large distance ($>5\arcmin$) with respect to the PSZ1 centre. In addition, the `Non detections' (ND) 
are, in practice, areas with an R compatible with the field galaxy count level, where no galaxy 
concentration is detected. Note that some SZ targets classified as ND report a galaxy system. In 
these cases, we want to enhance the existence of a galaxy system, but with too poor, showing 
an $R$ value to low ($R<15$), or even very low $sigma_v$, to be considered actual SZ counterpart 
under the restrictions imposed above.

In the majority of the cases, the analysis of the validation process following the confirmation 
criteria was unequivocal. However, in some fields, we found two or more separate systems, 
for which we obtained multiple photometric redshifts, which correspond to projection effects 
along the line of sight. These are named multiple detections. In addition, we detect clusters with multiple
clumps of galaxies assumed to be substructure (subclusters) of a single system. All these 
particular cases were investigated in detail by combining spectroscopic, photometric, and RGB 
images in order to disentangle the most probable counterpart for each SZ source.

Following this scheme, we consider `confirmed' optical counterparts those clusters classified 
with {\tt flag}=1 and 2, and `unconfirmed' targets, those associated with {\tt flag}=3 and 
an ND label.

\section{Results}
\label{sec:results}

Table ~\ref{tab:inpsz1} lists the 115 clusters from the PSZ1 catalogue explored in our optical 
follow-up. The first column shows the index number in the PSZ1 list. 
The second column lists the corresponding named in the PSZ1 catalogue. Column 3 is
the SZ significance reported in the PSZ1 catalogue. Columns 4, 5, and 6 are the J2000 equatorial 
coordinates corresponding to the most luminous cluster member (the BCG in most 
cases) and the distance between the optical and SZ centres. The multiple column 7 lists the mean 
spectroscopic redshift and the redshift of the BCG (if observed). Columns 8, 9, and 10 show the number 
of cluster members with spectroscopic measurements, the photometric redshift, and optical 
richness respectively. Column 11 lists the cluster classification following the flag scheme
described in section \ref{sec:criteria}, and column 10 adds some comments relative to other possible 
identifications or noteworthy features.
 
In accordance with in section \ref{sec:spec_data_subsec}, $z_{\rm spec,BCG}$ values 
present a mean error of $\Delta z=0.0004$. However, the error in $<z_{\rm spec}>$ 
is a bit higher, $\Delta z=0.001$, owing to the particular features of each cluster member sample, such
as the influence of the number of members considered, the 2D spatial distribution, presence of 
substructure, interlopers, etc. 

Following the confirmation criteria given above, we find that 53 SZ sources have reliable 
optical counterparts, 25 of them classified with {\tt Flag}$=1$ and 28 with {\tt Flag}$=2$. 
In addition, we classify 49 PSZ1 sources as `non-detections' (ND) and 13 as being weakly associated 
with the corresponding SZ source ({\tt Flag}$=3$). This means that a total of 62 SZ sources in
this sample remain unconfirmed. 

We report 56 spectroscopic redshifts, 30 of them found using the MOS mode of the TNG and GTC. 
Three of these SZ sources show multiple optical counterparts. The physical magnitudes associated
with these cluster counterparts, such as radial velocities of cluster members, velocity dispersions, 
and dynamical masses, will be discussed in detail in a future paper \citep{ferra2018}.

In the following subsections, we analyse the precision of the SZ coordinates and the agreement
between photometric and spectroscopic redshifts, and we discuss the nature of some of the 
clusters, in particular those systems presenting multiple optical counterparts and fossil 
systems. Finally, we study the presence of galactic dust as the most likely source of 
contamination in non-detections of SZ targets.


\subsection{Precision of SZ coordinates and redshifts}
\label{distances}

\begin{figure}[ht!]
\centering
\includegraphics[width=\columnwidth]{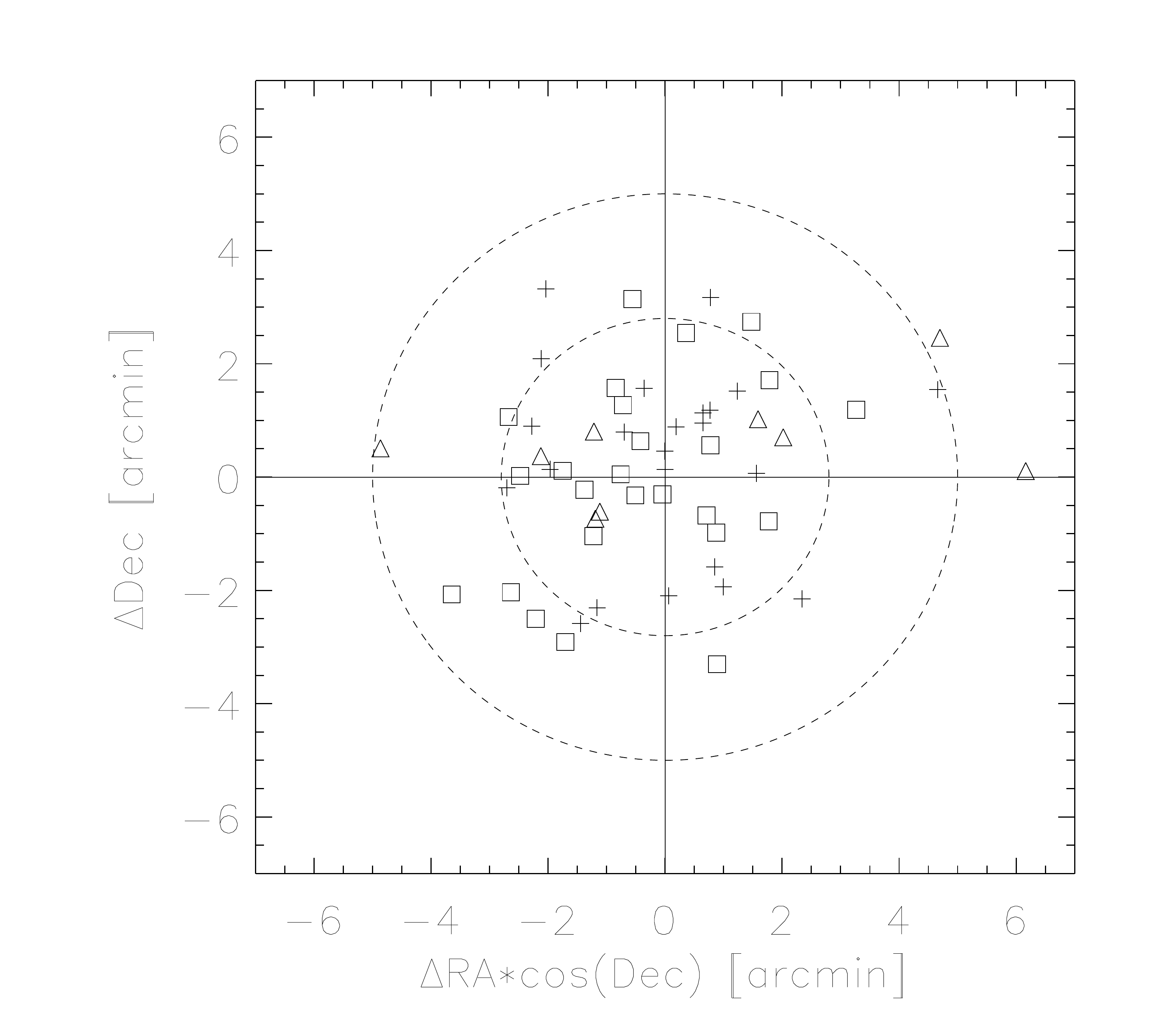}
\caption{Relative positions beween optical counterpart centres and nominal PSZ1 
coordinates for clusters in Table \ref{tab:inpsz1} flagged as `1' (crosses), 
`2' (squares), and `3' (triangles). Only optical counterparts with multiple 
detections are excluded. Inner dashed line show the region (of $2\parcm8$ radius)
enclosing 68\% of the confirmed clusters (flagged as `1' and `2'). External 
dashed line corresponds to 2.5 times the beam size ($5\arcm$) of \Planck\ SZ 
detections.}
\label{fig:offsets}
\end{figure}

Figure~\ref{fig:offsets} shows the spatial distribution of the optical centres of clusters 
relative to the PSZ1 coordinates. Multiple optical counterparts have been
excluded from this analysis. For these particular cases, it is difficult to determine
a single optical centre and how individual clusters contribute to the SZ emission. 
Twenty-five clusters in our sample containing 54 actual optical counterparts are closer than
$2\arcm$ to the \Planck\ SZ coordinate. In fact, 68\% of the clusters are enclosed 
within $2\parcm8$, which is in agreement with the value of $2\arcm$ 
predicted for the whole PSZ1 and found in the REFLEX II \Planck\ SZ sources.

\begin{landscape}
\begin{table}
\begingroup
\caption{Clusters candidates from the PSZ1 catalogue studied in this work.}
\label{tab:inpsz1}
\nointerlineskip
\vskip -3mm
\scriptsize
   \newdimen\digitwidth
   \setbox0=\hbox{\rm 0}
   \digitwidth=\wd0
   \catcode`*=\active
   \def*{\kern\digitwidth}
   \newdimen\signwidth
   \setbox0=\hbox{+}
   \signwidth=\wd0
   \catcode`!=\active
   \def!{\kern\signwidth}
\halign{\hfil#\hfil\tabskip=3em&
   \hbox to 0.9in{#}\tabskip=2em&
   \hfil#\hfil&
  \hfil#\hfil\tabskip=1em&
   \hfil#\hfil\tabskip=2em& 
   \hfil#\hfil&
   \hfil#\hfil&
   \hfil#\hfil&
   \hfil#\hfil&
   \hfil#\hfil&
   \hfil#\hfil&
   #\hfil\tabskip=0pt\cr
\omit&\omit&\omit&\multispan2\hfil Position (J2000)\hfil&\omit\cr
\noalign{\vskip -3pt}
\omit&\omit&\omit&\multispan2\hrulefill&\omit\cr
ID\rlap{$^{\rm 1}$}&\omit\hfil \Planck\ Name\hfil& SZ $\ S/N$ & R. A.&Decl.& Dist. ($\arcm$) & $<z_{\rm spec}>$~;~$z_{\rm spec,BCG}$&$N_{\rm spec}$&$z_{\rm phot}$&R&
{\tt Flag} &\omit\hfil Notes\hfil\cr
\noalign{\vskip 3pt\hrule\vskip 5pt}
***4  	                 & PSZ1 G001.00$+$25.71 & 6.04 &    $-$	     &     $-$	    & $-$  &      $-$	      & $-$ &  $-$          & $-$	  & ND &  \cr
**56	                 & PSZ1 G021.88$+$17.75 & 5.64 & 17 28 16.01 & -01 22 58.04 & 2.58 & 0.646 ~;~ 0.6488 & 10  & 0.60$\pm$0.07 & *44$\pm$6.6 & *1 &  \cr
**67	                 & PSZ1 G027.31$+$23.79 & 5.10 &    $-$      &	   $-$	    & $-$  &	  $-$	      & $-$ &  $-$	    & $-$	  & ND &  \cr
**68	                 & PSZ1 G027.75$+$15.41 & 4.67 &    $-$      &	   $-$	    & $-$  &	  $-$	      & $-$ &  $-$	    & $-$	  & ND &  \cr 
**69	                 & PSZ1 G027.95$+$15.63 & 4.77 &    $-$      &	   $-$	    & $-$  &	  $-$	      & $-$ &  $-$	    & $-$	  & ND &  \cr
**70-A                   & PSZ1 G028.01$+$25.46 & 5.98 & 17 11 54.45 & +07 23 13.37 & 5.16 & 0.658 ~;~ 0.6641 & *9  & 0.61$\pm$0.06 & *17$\pm$4.1 & *1 &  Liu+15 report a different counterpart \cr 
**70-B                   &			&      & 17 11 21.53 & +07 19 23.17 & 6.01 &      $-$	      & $-$ & 0.60$\pm$0.06 & *18$\pm$4.2 & *3 &  \cr 
**72	                 & PSZ1 G028.15$-$08.63 & 5.07 &    $-$      &	   $-$	    & $-$  &	  $-$	      & $-$ &  $-$	    & $-$	  & ND &  \cr
**78	                 & PSZ1 G029.79$-$17.37 & 6.59 &    $-$      &	   $-$	    & $-$  &	  $-$	      & $-$ &  $-$	    & $-$	  & ND &  \cr  
**79	                 & PSZ1 G030.21$-$16.91 & 5.34 &    $-$      &	   $-$	    & $-$  &	  $-$	      & $-$ &  $-$	    & $-$	  & ND &  \cr
**80			 & PSZ1 G030.70$+$09.47 & 4.66 & 18 13 58.80 & +02 22 32.30 & 4.90 & 0.052 ~;~ 0.0532 & 17  & 0.11$\pm$0.02 & *62$\pm$7.9 & *1 &  \cr
**82                     & PSZ1 G030.98$+$22.43 & 4.98 & 17 27 59.37 & +08 23 50.10 & 6.10 & 0.424 ~;~ 0.4236 & *2  & 0.50$\pm$0.06 & *17$\pm$4.1 & *3 &  \cr  
**84			 & PSZ1 G031.41$+$28.75 & 4.84 & 17 04 58.60 & +11 27 01.00 & 1.27 & 0.230 ~;~ 0.2304 & *4  & 0.24$\pm$0.03 & *16$\pm$4.0 & *3 &  vdB+16 report no counterpart \cr
**88			 & PSZ1 G032.15$-$14.93 & 8.21 & 19 43 11.20 & -07 24 56.25 & 4.89 & 0.377 ~;~ 0.3775 & 10  & 0.43$\pm$0.04 & *51$\pm$7.1 & *3 &  Substructured \cr
**90			 & PSZ1 G032.76$+$42.33 & 4.68 & 16 15 05.78 & +17 46 52.20 & 0.48 & 0.844 ~;~ 0.8431 & *8  & 0.67$\pm$0.05 & *39$\pm$6.2 & *1 &  \cr
**91	                 & PSZ1 G033.33$-$17.54 & 5.78 & 19 54 59.67 & -07 30 34.70 & 2.57 &      $-$	      & $-$ & 0.33$\pm$0.04 & *23$\pm$4.8 & *2 &  \cr
**98	                 & PSZ1 G036.02$+$12.21 & 4.70 &    $-$      &	   $-$	    & $-$  &	  $-$	      & $-$ &  $-$	    & $-$	  & ND &  \cr
**99	                 & PSZ1 G036.09$-$17.40 & 5.35 &    $-$      &	   $-$	    & $-$  &	  $-$	      & $-$ &  $-$	    & $-$	  & ND &  \cr
*101	                 & PSZ1 G037.22$-$16.24 & 5.02 &    $-$      &	   $-$	    & $-$  &	  $-$	      & $-$ &  $-$	    & $-$	  & ND &  \cr
*121                     & PSZ1 G042.96$+$19.11 & 4.70 & 17 58 55.06 & +17 13 33.40 & 2.45 & 0.499 ~;~ 0.4993 & *8  & 0.53$\pm$0.05 & *45$\pm$6.7 & *1 &  Substructured \cr
*126                     & PSZ1 G044.82$-$31.66 & 4.67 & 21 04 46.58 & -04 45 44.50 & 2.14 & 0.221 ~;~ 0.2213 & *7  & 0.21$\pm$0.02 & *23$\pm$4.8 & *3 &  Liu+15 cluster 126 \cr  
*127	                 & PSZ1 G044.83$+$10.02 & 7.27 &    $-$      &	   $-$	    & $-$  &	  $-$	      & $-$ &  $-$	    & $-$	  & ND &  \cr
*133                     & PSZ1 G045.54$+$16.26 & 4.67 & 18 14 13.31 & +18 17 03.61 & 2.18 & 0.206 ~;~ 0.2058 & 21  & 0.22$\pm$0.02 & *49$\pm$7.0 & *1 &  \cr	   
*136	                 & PSZ1 G046.02$-$09.13 & 4.74 &    $-$      &	   $-$	    & $-$  &	  $-$	      & $-$ &  $-$	    & $-$	  & ND &  \cr
*139	                 & PSZ1 G046.35$-$06.83 & 4.80 &    $-$      &     $-$      & $-$  &      $-$	      & $-$ &  $-$	    & $-$	  & ND &  \cr
*143			 & PSZ1 G047.44$+$37.39 & 4.92 & 16 50 20.41 & +26 58 21.40 & 3.18 & 0.230 ~;~ 0.2318 & 10  & 0.23$\pm$0.02 & *17$\pm$4.1 & *1 &  \cr
*144	                 & PSZ1 G047.53$+$08.55 & 5.82 &    $-$      &	   $-$	    & $-$  &	  $-$	      & $-$ &  $-$	    & $-$	  & ND &  \cr
*158	                 & PSZ1 G050.01$-$16.88 & 5.29 &    $-$      &	   $-$	    & $-$  &	  $-$	      & $-$ &  $-$	    & $-$	  & ND &  \cr
*163                     & PSZ1 G052.93$+$10.44 & 4.91 & 18 49 11.97 & +22 26 39.36 & 4.09 & 0.219 ~;~ 0.2194 & 10  & 0.24$\pm$0.03 & *12$\pm$3.5 & ND &  \cr
*165	                 & PSZ1 G053.50$+$09.56 & 5.11 &    $-$      &	   $-$	    & $-$  &	  $-$	      & $-$ &  $-$	    & $-$	  & ND &  vdB+16 report no counterpart \cr
*170			 & PSZ1 G054.59$-$18.18 & 5.33 & 20 36 38.70 & +09 47 53.70 & 4.40 & 0.446 ~;~ 0.4492 & *4  & 0.38$\pm$0.04 & *14$\pm$3.7 & ND &  \cr
*176\rlap{$^{\rm a}$}    & PSZ1 G055.83$-$41.64 & 5.72 & 21 56 41.02 & -02 32 20.87 & 8.29 &      $-$	      & $-$ & 0.63$\pm$0.04 & *16$\pm$4.0 & *3 &  \cr
*179			 & PSZ1 G056.76$-$11.60 & 6.56 & 20 18 57.20 & +15 07 18.3  & 1.41 & 0.123 ~;~ 0.1217 & 21  &  $-$	    & *66$\pm$8.1 & *1 &  \cr
*184			 & PSZ1 G057.42$-$10.77 & 4.76 & 20 17 25.72 & +16 03 27.30 & 2.10 & 0.136 ~;~ 0.1360 & 25  & 0.06$\pm$0.02 & *34$\pm$5.8 & *1 &  \cr
*192	                 & PSZ1 G058.45$-$33.47 & 4.83 &    $-$      &     $-$      & $-$  &      $-$	      & $-$ &  $-$	    & $-$	  & ND &  \cr
*213			 & PSZ1 G065.13$+$57.53 & 4.70 & 15 16 02.04 & +39 44 26.40 & 2.71 & 0.684 ~;~ 0.6857 & 20  & 0.69$\pm$0.04 & *78$\pm$8.8 & *1 &  Substructured \cr
*239\rlap{$^{\rm a}$}	 & PSZ1 G071.64$-$42.76 & 8.82 & 22 30 50.00 & +05 39 16.72 & 1.88 &      $-$	      & $-$ & 0.69$\pm$0.08 & **7$\pm$2.6 & ND &  vdB+16 invalidate this source \cr
*251-A                   & PSZ1 G075.29$+$26.66 & 5.17 & 18 08 44.26 & +47 41 09.22 & 5.85 & 0.281 ~;~ 0.2816 & 10  & 0.21$\pm$0.03 & *16$\pm$4.0 & *1 &  \cr
*251-B                   &			&      & 18 09 09.02 & +47 49 01.11 & 5.98 &      $-$	      & $-$ & 0.25$\pm$0.03 & *17$\pm$4.1 & *2 &  \cr
*257\rlap{$^{\rm a}$}	 & PSZ1 G078.39$+$46.13 & 4.90 & 16 09 01.45 & +50 05 11.32 & 7.46 & 0.400 ~;~ 0.3999 & *4  & 0.41$\pm$0.04 & *38$\pm$6.2 & *3 &  Liu+15 cluster 257 \cr
*261	                 & PSZ1 G079.88$+$14.97 & 4.71 & 19 23 12.07 & +48 16 13.25 & 0.91 & 0.101 ~;~ 0.1020 & 12  & 0.07$\pm$0.02 & *56$\pm$7.5 & *1 &  \cr
*271\rlap{$^{\rm a}$}	 & PSZ1 G081.56$+$31.03 & 4.71 & 17 45 54.62 & +53 48 45.96 & 3.33 &      $-$	      & $-$ & 0.79$\pm$0.06 & *22$\pm$4.7 & *2 &  vdB+16 invalidate this source \cr
*279                     & PSZ1 G084.04$+$58.75 & 4.84 & 14 49 00.90 & +48 33 24.00 & 2.98 & 0.735 ~;~ 0.7302 & *5  & 0.70$\pm$0.06 & *58$\pm$7.6 & *1 &  \cr 
*289                     & PSZ1 G085.71$+$10.67 & 5.35 & 20 03 13.30 & +51 20 51.00 & 1.56 & 0.084 ~;~ 0.0804 & 12  & 0.06$\pm$0.02 & *70$\pm$8.4 & *1 &  \cr
*305                     & PSZ1 G090.14$-$49.71 & 4.82 &    $-$      &     $-$      & $-$  &      $-$	      & $-$ &  $-$	    & $-$	  & ND &  Liu+15 report a counterpart at $z_{\rm phot}$=0.207 \cr
*306-A                   & PSZ1 G090.48$+$46.89 & 4.90 & 15 45 18.76 & +57 43 37.59 & 3.82 &      $-$	      & $-$ & 0.54$\pm$0.04 & *15$\pm$3.9 & *3 &  \cr
*306-B                   &			&      & 15 44 07.40 & +57 46 43.20 & 9.26 & 0.676 ~;~ 0.6725 & *6  & 0.70$\pm$0.05 & *22$\pm$4.7 & *3 &  \cr
*310\rlap{$^{\rm a}$}	 & PSZ1 G091.14$-$38.73 & 4.79 & 23 09 10.67 & +17 47 38.29 & 3.19 & 0.105 ~;~ 0.1048 & *1  & 0.10$\pm$0.02 & *42$\pm$6.5 & *2 &  \cr
*314                     & PSZ1 G091.93$+$35.48 & 5.18 & 17 09 52.64 & +62 22 07.67 & 2.16 & 0.276 ~;~ 0.2755 & 14  & 0.26$\pm$0.02 & *30$\pm$5.5 & *3 &  Liu+15 cluster 314 \cr  
*318     	         & PSZ1 G092.46$-$35.25 & 5.47 &    $-$      &     $-$      & $-$  &       $-$	      & $-$ &  $-$	    & $-$	  & ND &  \cr
*320      	         & PSZ1 G093.04$-$32.38 & 5.69 & 23 02 15.07 & +24 03 50.50 & 7.48 & 0.512 ~;~ 0.5104 & *5  & 0.55$\pm$0.05 & *42$\pm$6.5 & *3 &  \cr
*331      	         & PSZ1 G094.95$-$36.72 & 4.79 & 23 16 25.10 & +20 57 23.90 & 1.79 &       $-$	      & $-$ & 0.50$\pm$0.05 & *19$\pm$4.4 & *2 &  \cr
*336            	 & PSZ1 G096.44$-$10.40 & 6.55 & 22 20 12.95 & +44 26 16.27 & 8.17 & 0.195 ~;~ 0.1948 & *4  & 0.22$\pm$0.03 & *14$\pm$3.7 & ND &  \cr
*348                     & PSZ1 G098.67$-$07.03 & 4.99 &    $-$      &     $-$      & $-$  &       $-$	      & $-$ &  $-$	    & $-$	  & ND &  \cr
*349	                 & PSZ1 G098.85$-$07.27 & 4.87 &    $-$      &	   $-$	    & $-$  &	   $-$	      & $-$ &  $-$	    & $-$	  & ND &  \cr
*356	                 & PSZ1 G099.63$+$14.85 & 4.72 &    $-$      &	   $-$	    & $-$  &	   $-$	      & $-$ &  $-$	    & $-$	  & ND &  \cr
*361\rlap{$^{\rm a}$}	 & PSZ1 G100.46$-$61.45 & 4.71 & 00 09 15.84 & -00 26 58.94 & 1.61 & 0.256 ~;~ 0.2561 & *1  & 0.28$\pm$0.02 & *18$\pm$4.2 & *2 &  \cr
*369                     & PSZ1 G102.97$-$04.77 & 5.64 & 22 34 47.08 & +52 42 55.56 & 0.31 &       $-$	      & $-$ & 0.52$\pm$0.04 & *27$\pm$5.2 & *2 &  \cr
\noalign{\vskip 3pt\hrule\vskip 3pt}}
\endgroup
\end{table}
\end{landscape}

\addtocounter{table}{-1}
\begin{landscape}
\begin{table}
\begingroup
\caption{Continue.}
\nointerlineskip
\vskip -3mm
\scriptsize
   \newdimen\digitwidth
   \setbox0=\hbox{\rm 0}
   \digitwidth=\wd0
   \catcode`*=\active
   \def*{\kern\digitwidth}
   \newdimen\signwidth
   \setbox0=\hbox{+}
   \signwidth=\wd0
   \catcode`!=\active
   \def!{\kern\signwidth}
\halign{\hfil#\hfil\tabskip=3em&
   \hbox to 0.9in{#}\tabskip=2em&
   \hfil#\hfil&
   \hfil#\hfil\tabskip=1em&
   \hfil#\hfil\tabskip=2em& 
   \hfil#\hfil&
   \hfil#\hfil&
   \hfil#\hfil&
   \hfil#\hfil&
   \hfil#\hfil&
   \hfil#\hfil&
   #\hfil\tabskip=0pt\cr
\omit&\omit&\omit&\multispan2\hfil Position (J2000)\hfil&\omit\cr
\noalign{\vskip -3pt}
\omit&\omit&\omit&\multispan2\hrulefill&\omit\cr
ID\rlap{$^{\rm 1}$}&\omit\hfil \Planck\ Name\hfil& SZ $\ S/N$ &R. A.&Decl.& Dist. ($\arcm$) & $<z_{\rm spec}>$~;~$z_{\rm spec,BCG}$&$N_{\rm spec}$&$z_{\rm phot}$&R&\omit\hfil Notes\hfil\cr
\noalign{\vskip 3pt\hrule\vskip 5pt}
*370 	                 & PSZ1 G103.16$-$14.95 & 5.08 & 23 04 00.34 & +43 46 21.40 & 1.99 &       $-$	      & $-$ & 0.27$\pm$0.03 & *13$\pm$3.6 & ND &  \cr
*375            	 & PSZ1 G103.94$+$25.81 & 4.78 & 18 52 09.50 & +72 59 33.12 & 1.96 & 0.077 ~;~ 0.0777 & 17  & 0.06$\pm$0.02 & *66$\pm$8.1 & *1 &  Liu cluster 375. Fossil system \cr
*377			 & PSZ1 G104.78$+$40.45 & 4.92 & 15 46 29.88 & +69 57 59.11 & 2.16 & 0.837 ~;~ 0.8373 & 10  & 0.70$\pm$0.06 & *40$\pm$6.3 & *1 &  \cr
*382             	 & PSZ1 G106.07$-$17.42 & 4.73 & 23 23 20.46 & +42 33 15.32 & 1.89 & 0.817 ~;~ 0.8144 & 10  & 0.65$\pm$0.07 & *21$\pm$4.6 & *3 &  \cr
*385                     & PSZ1 G106.49$-$10.43 & 5.40 & 23 10 58.60 & +49 12 28.21 & 3.80 &       $-$	      & $-$ & 0.49$\pm$0.05 & **9$\pm$3.0 & ND &  \cr
*387	                 & PSZ1 G106.81$-$36.44 & 4.66 &    $-$      &	   $-$	    & $-$  &	   $-$	      & $-$ &  $-$	    & $-$	  & ND &  \cr
*393	                 & PSZ1 G108.13$-$09.21 & 5.39 &    $-$      &	   $-$	    & $-$  &	   $-$	      & $-$ &  $-$	    & $-$	  & ND &  \cr
*395			 & PSZ1 G108.26$+$48.66 & 4.88 & 14 27 04.57 & +65 39 46.94 & 1.06 & 0.671 ~;~ 0.6752 & 29  & 0.69$\pm$0.06 & *34$\pm$5.8 & *1 &  Substructured \cr
*397                     & PSZ1 G108.90$-$52.04 & 6.89 & 00 16 26.74 & +09 53 53.60 & 2.88 & 0.461 ~;~ 0.4620 & *4  & 0.54$\pm$0.04 & *23$\pm$4.8 & *2 &  \cr
*412                     & PSZ1 G112.60$-$39.30 & 4.76 &    $-$      &     $-$      & $-$  &       $-$	      & $-$ &  $-$	    & $-$	  & ND &  \cr
*420	                 & PSZ1 G114.81$-$11.80 & 4.85 & 00 01 14.70 & +50 16 32.16 & 2.48 & 0.228 ~;~ 0.2281 & *1  & 0.21$\pm$0.02 & *38$\pm$6.2 & *2 &  Inlcuded in RTT150 work \cr 
*421	                 & PSZ1 G114.98$+$19.10 & 5.19 &    $-$      &	   $-$	    & $-$  &	   $-$	      & $-$ &  $-$	    & $-$	  & ND &  \cr 
*436	                 & PSZ1 G118.61$+$10.55 & 4.96 & 23 53 24.56 & +72 59 00.33 & 3.11 &	   $-$	      & $-$ & 0.77$\pm$0.07 & *50$\pm$7.1 & *2 &  \cr
*444	                 & PSZ1 G121.27$+$23.08 & 4.72 &    $-$      &	   $-$	    & $-$  &	   $-$	      & $-$ &  $-$	    & $-$	  & ND &  \cr
*446	                 & PSZ1 G121.69$+$24.47 & 5.35 &    $-$      &	   $-$	    & $-$  &	   $-$	      & $-$ &  $-$	    & $-$	  & ND &  \cr
*449	                 & PSZ1 G123.37$+$25.34 & 5.90 & 01 45 31.18 & +88 12 16.91 & 2.48 &	   $-$	      & $-$ & 0.95$\pm$0.11 & *22$\pm$4.7 & *2 &  \cr 
*453	                 & PSZ1 G123.79$+$25.86 & 5.04 & 03 00 48.59 & +88 30 09.18 & 1.31 &	   $-$	      & $-$ & 0.54$\pm$0.04 & *48$\pm$6.9 & *2 &  \cr
*456	                 & PSZ1 G124.64$+$29.38 & 4.89 & 10 35 55.45 & +87 16 34.35 & 1.39 &	   $-$	      & $-$ & 0.35$\pm$0.03 & *36$\pm$6.0 & *2 &  \cr
*458-A\rlap{$^{\rm a}$}	 & PSZ1 G125.54$-$56.25 & 4.86 & 00 57 09.25 & +06 35 05.92 & 2.04 & 0.548 ~;~ 0.5475 & *1  & 0.59$\pm$0.04 & *26$\pm$5.1 & *2 &  \cr
*458-B\rlap{$^{\rm a}$}	 &			&      & 00 57 08.83 & +06 34 06.14 & 2.49 & 0.168 ~;~ 0.1676 & *1  & 0.17$\pm$0.02 & *32$\pm$5.7 & *2 &  \cr
*458-C\rlap{$^{\rm a}$}	 &			&      & 00 57 22.23 & +06 38 27.27 & 3.29 & 0.296 ~;~ 0.2961 & *1  & 0.27$\pm$0.03 & *33$\pm$5.7 & *2 &  \cr
*462\rlap{$^{\rm a}$}    & PSZ1 G126.44$-$70.36 &      & 00 55 57.36 & -07 33 42.46 & 4.20 & 0.325 ~;~ 0.3246 & *3  & 0.33$\pm$0.04 & *21$\pm$4.6 & *2 &  \cr  
*465	                 & PSZ1 G127.36$-$10.69 & 5.58 &    $-$      &	   $-$	    & $-$  &	   $-$	      & $-$ &  $-$	    & $-$	  & ND &  \cr
*468	                 & PSZ1 G128.75$-$17.97 & 4.95 &    $-$      &	   $-$	    & $-$  &	   $-$	      & $-$ &  $-$	    & $-$	  & ND &  \cr
*476                     & PSZ1 G133.50$-$46.77 & 4.91 & 01 21 25.03 & +15 30 56.84 & 1.38 &       $-$	      & $-$ & 0.68$\pm$0.09 & *14$\pm$3.7 & ND &  \cr  
*478        	         & PSZ1 G134.08$-$44.61 & 5.42 &    $-$      &     $-$      & $-$  &       $-$	      & $-$ &  $-$	    & $-$	  & ND &  \cr
*479                     & PSZ1 G134.31$-$06.57 & 4.81 & 02 10 25.10 & +54 34 09.80 & 1.61 & 0.333 ~;~ 0.3341 & 20  & 0.35$\pm$0.03 & *64$\pm$8.0 & *1 &  \cr
*490                     & PSZ1 G135.92$+$76.21	& 5.22 & 12 35 46.89 & +40 27 55.53 & 6.29 & 0.406 ~;~   $-$  & *9  & 0.39$\pm$0.04 & *14$\pm$3.7 & ND &  \cr
*492                     & PSZ1 G136.37$-$44.50 & 4.90 &    $-$      &     $-$      & $-$  &	   $-$        & $-$ &  $-$	    & $-$	  & ND &  \cr
*496                     & PSZ1 G137.51$-$10.01 & 5.33 & 02 22 53.51 & +50 14 40.11 & 1.93 &	   $-$        & $-$ & 0.14$\pm$0.02 & *67$\pm$8.2 & *2 &  \cr
*497                     & PSZ1 G137.56$+$53.88 & 5.73 &    $-$      &     $-$      & $-$  & 	   $-$        & $-$ &  $-$          & $-$         & ND &  \cr
*504  	                 & PSZ1 G140.10$+$50.09 & 4.90 & 11 11 30.47 & +63 35 16.76 & 5.33 &       $-$	      & $-$ & 0.70$\pm$0.05 & *26$\pm$5.1 & *3 &  \cr
*509			 & PSZ1 G142.17$+$37.28 & 5.79 & 09 19 05.16 & +70 55 11.40 & 3.26 & 0.240 ~;~ 0.2393 & *6  & 0.23$\pm$0.03 & *74$\pm$8.6 & *1 &  Liu+15 report a $z_{\rm phot}$=0.28 \cr
*511                     & PSZ1 G142.38$+$22.82 & 5.81 & 06 13 49.84 & +71 52 54.78 & 0.96 & 0.394 ~;~ 0.3927 & *2  & 0.34$\pm$0.05 & *20$\pm$4.5 & *2 &  \cr  
*529	                 & PSZ1 G148.20$+$23.49 & 8.40 & 06 37 54.60 & +66 51 06.20 & 3.42 & 0.098 ~;~ 0.0980 & *2  &  $-$	    & *35$\pm$5.9 & *2 &  Liu+15 cluster 529 \cr
*534                     & PSZ1 G150.33$-$20.04 & 4.75 & 02 59 31.65 & +35 56 30.47 & 11.7 &       $-$	      & $-$ & 0.07$\pm$0.01 & *15$\pm$3.9 & ND &  \cr  	  
*539                     & PSZ1 G151.80$-$48.06 & 5.59 & 02 08 07.55 & +10 27 17.54 & 0.98 & 0.201 ~;~ 0.1999 & *2  &  $-$	    & *23$\pm$4.8 & *2 &  ACO 307 \cr
*549                     & PSZ1 G157.07$-$33.66 & 4.90 & 02 51 35.96 & +21 07 05.30 & 0.60 &	   $-$        & $-$ & 0.62$\pm$0.05 & *21$\pm$4.6 & *2 &  vdB+16 invalidate this source \cr
*551	                 & PSZ1 G157.44$+$30.34 & 7.54 & 07 48 54.35 & +59 42 05.77 & 1.46 & 0.407 ~;~ 0.4046 & *4  &  $-$	    & *18$\pm$4.2 & *2 &  [ATZ98] B100 \cr
*564	                 & PSZ1 G162.30$-$26.92 & 6.56 & 03 24 19.02 & +23 57 49.82 & 3.37 & 0.391 ~;~ 0.3917 & *3  & 0.34$\pm$0.03 & *44$\pm$6.6 & *2 &  \cr
*586-A                   & PSZ1 G169.80$+$26.10 & 5.32 & 07 30 32.02 & +48 17 39.05 & 4.21 &	   $-$        & $-$ & 0.73$\pm$0.07 & *36$\pm$6.0 & *2 &  \cr
*586-B                   &			&      & 07 30 29.09 & +48 20 39.15 & 3.53 &	   $-$        & $-$ & 0.83$\pm$0.10 & *19$\pm$4.6 & *2 &  \cr
*605	                 & PSZ1 G178.10$+$18.58 & 5.01 & 07 01 31.33 & +38 52 48.60 & 9.28 &	   $-$        & $-$ & 0.38$\pm$0.03 & *12$\pm$3.5 & ND &  \cr
*612             	 & PSZ1 G181.21$-$30.73 & 5.87 & 04 02 56.76 & +09 44 29.10 & 0.76 & 0.540 ~;~ 0.5406 & *3  & 0.41$\pm$0.04 & *26$\pm$5.1 & *2 &  Liu+15 cluster 612 \cr
*618-A	                 & PSZ1 G183.26$+$12.25 & 5.43 & 06 43 09.84 & +31 50 55.47 & 3.44 & 0.638 ~;~ 0.6352 & *2  & 0.62$\pm$0.05 & *25$\pm$5.0 & *2 &  See Fig.~\ref{fig:cmd} \cr
*618-B	                 &			&      & 06 42 58.24 & +31 45 01.07 & 4.12 &       $-$	      & $-$ & 0.27$\pm$0.03 & *23$\pm$4.8 & *2 &  \cr
*624                     & PSZ1 G185.42$-$32.03 & 6.15 & 04 07 50.12 & +06 07 06.29 & 1.76 &       $-$	      & $-$ & 0.08$\pm$0.02 & *43$\pm$6.6 & *2 &  \cr
*626                     & PSZ1 G185.93$-$31.21 & 5.90 & 04 11 52.31 & +06 17 11.80 & 3.47 & 0.094 ~;~ 0.0947 & *2  & 0.11$\pm$0.02 & *22$\pm$4.7 & *2 &  Fossil system \cr
*634                     & PSZ1 G188.36$-$35.00 & 5.28 & 04 04 18.02 & +02 23 55.46 & 1.30 & 0.275 ~;~ 0.2723 & *4  &  $-$	    & *29$\pm$5.4 & *1 &  ZwCl 0401.8+0219 \cr
*641                     & PSZ1 G189.82$-$37.25 & 6.99 &    $-$      &     $-$      & $-$  &       $-$	      & $-$ &  $-$          & $-$	  & ND &  \cr
*653	                 & PSZ1 G194.74$-$10.10 & 5.19 & 05 41 05.73 & +11 10 05.32 & 0.76 &       $-$	      & $-$ & 0.75$\pm$0.10 & *30$\pm$5.5 & *2 &  \cr
*682                     & PSZ1 G206.45$+$13.89 & 5.90 & 07 29 51.23 & +11 56 30.89 & 1.97 & 0.406 ~;~ 0.4055 & 45  & 0.44$\pm$0.04 & 127$\pm$11.2& *1 &  Gravitational arc \cr
*684 	                 & PSZ1 G206.64$-$21.17 & 6.62 &    $-$      &     $-$      & $-$  &       $-$	      & $-$ & $-$           & $-$	  & ND &  \cr
*713	                 & PSZ1 G216.27$+$10.10 & 5.16 & 07 33 20.03 & +01 36 36.06 & 3.14 &	   $-$        & $-$ & 0.15$\pm$0.02 & *31$\pm$5.6 & *2 &  \cr
*723	                 & PSZ1 G218.54$+$13.26 & 5.24 & 07 48 51.66 & +01 06 39.88 & 1.80 & 0.266 ~;~ 0.2682 & 16  & 0.24$\pm$0.03 & *77$\pm$8.8 & *1 &  \cr
*752	                 & PSZ1 G224.82$+$13.62 & 5.51 & 08 01 41.61 & -04 03 46.23 & 0.14 & 0.274 ~;~ 0.2759 & 28  & 0.25$\pm$0.03 & 132$\pm$11.5& *1 &  \cr
*827                     & PSZ1 G244.48$+$34.06 & 8.14 & 09 49 46.96 & -07 30 12.50 & 1.40 & 0.135 ~;~ 0.1342 & *7  & 0.15$\pm$0.02 & *23$\pm$4.8 & *3 &  \cr 
*992                     & PSZ1 G286.25$+$62.68 & 5.52 & 12 21 05.35 & +00 48 22.29 & 1.46 & 0.211 ~;~ 0.2107 & 11  & 0.18$\pm$0.03 & *48$\pm$6.9 & *3 &  \cr
\noalign{\vskip 3pt\hrule\vskip 3pt}}
\endgroup
\end{table}
\end{landscape}

\addtocounter{table}{-1}
\begin{landscape}
\begin{table}
\begingroup
\caption{Continue.}
\nointerlineskip
\vskip -3mm
\scriptsize
   \newdimen\digitwidth
   \setbox0=\hbox{\rm 0}
   \digitwidth=\wd0
   \catcode`*=\active
   \def*{\kern\digitwidth}
   \newdimen\signwidth
   \setbox0=\hbox{+}
   \signwidth=\wd0
   \catcode`!=\active
   \def!{\kern\signwidth}
\halign{\hfil#\hfil\tabskip=3em&
   \hbox to 0.9in{#}\tabskip=2em&
   \hfil#\hfil&
   \hfil#\hfil\tabskip=1em&
   \hfil#\hfil\tabskip=2em& 
   \hfil#\hfil&
   \hfil#\hfil&
   \hfil#\hfil&
   \hfil#\hfil&
   \hfil#\hfil&
   \hfil#\hfil&
   #\hfil\tabskip=0pt\cr
\omit&\omit&\omit&\multispan2\hfil Position (J2000)\hfil&\omit\cr
\noalign{\vskip -3pt}
\omit&\omit&\omit&\multispan2\hrulefill&\omit\cr
ID\rlap{$^{\rm 1}$}&\omit\hfil \Planck\ Name\hfil& SZ $\ S/N$ &R. A.&Decl.& Dist. ($\arcm$) & $<z_{\rm spec}>$~;~$z_{\rm spec,BCG}$&$N_{\rm spec}$&$z_{\rm phot}$&R&\omit\hfil Notes\hfil\cr
\noalign{\vskip 3pt\hrule\vskip 5pt}
1122                     & PSZ1 G318.61$+$83.80 & 6.93 &    $-$      &     $-$      & $-$  &       $-$	      & $-$ &  $-$	    & $-$	  & ND &  \cr
1159\rlap{$^{\rm a}$}    & PSZ1 G332.30$+$72.17 & 4.76 & 13 26 33.96 & +11 18 06.51 & 2.96 & 0.089 ~;~ 0.0898 & 30  & 0.06$\pm$0.02 & *64$\pm$8.0 & *1 &  Liu+15 cluster 1159 \cr
1189                     & PSZ1 G341.69$+$50.66 & 5.48 & 14 25 12.29 & -04 56 34.19 & 3.90 & 0.293 ~;~ 0.2913 & 31  & 0.25$\pm$0.03 & *35$\pm$5.9 & *1 &  Substructured. Liu+15 cluster 1189-A \cr
1198	                 & PSZ1 G345.81$+$42.38 & 4.80 &    $-$      &	  $-$	    & $-$  &	   $-$	      & $-$ &  $-$	    & $-$	  & ND &  \cr  
1212	                 & PSZ1 G352.04$+$42.25 & 4.74 &    $-$      &	  $-$	    & $-$  &	   $-$	      & $-$ &  $-$	    & $-$	  & ND &  \cr
1221             	 & PSZ1 G357.43$+$30.60 & 5.47 & 15 54 53.89 & -12 13 16.75 & 4.45 &       $-$	      & $-$ & 0.13$\pm$0.02 & *13$\pm$3.6 & ND &  \cr
\noalign{\vskip 3pt\hrule\vskip 3pt}}
 \begin{tablenotes} 
  \tiny
  \item{\textit{$^{\rm 1}$ SZ targets identified with the ID followed by an ``A'', ``B'' or ``C'' label indicate the presence
  of multiple counterparts.}}
  \item{\textit{$^{\rm a}$ Photometric and/or spectroscopic redshift obtained from SDSS DR12 data.}}
 \end{tablenotes}
\endgroup
\end{table}
\end{landscape}

\begin{figure}[ht!]
\centering
\includegraphics[width=\columnwidth]{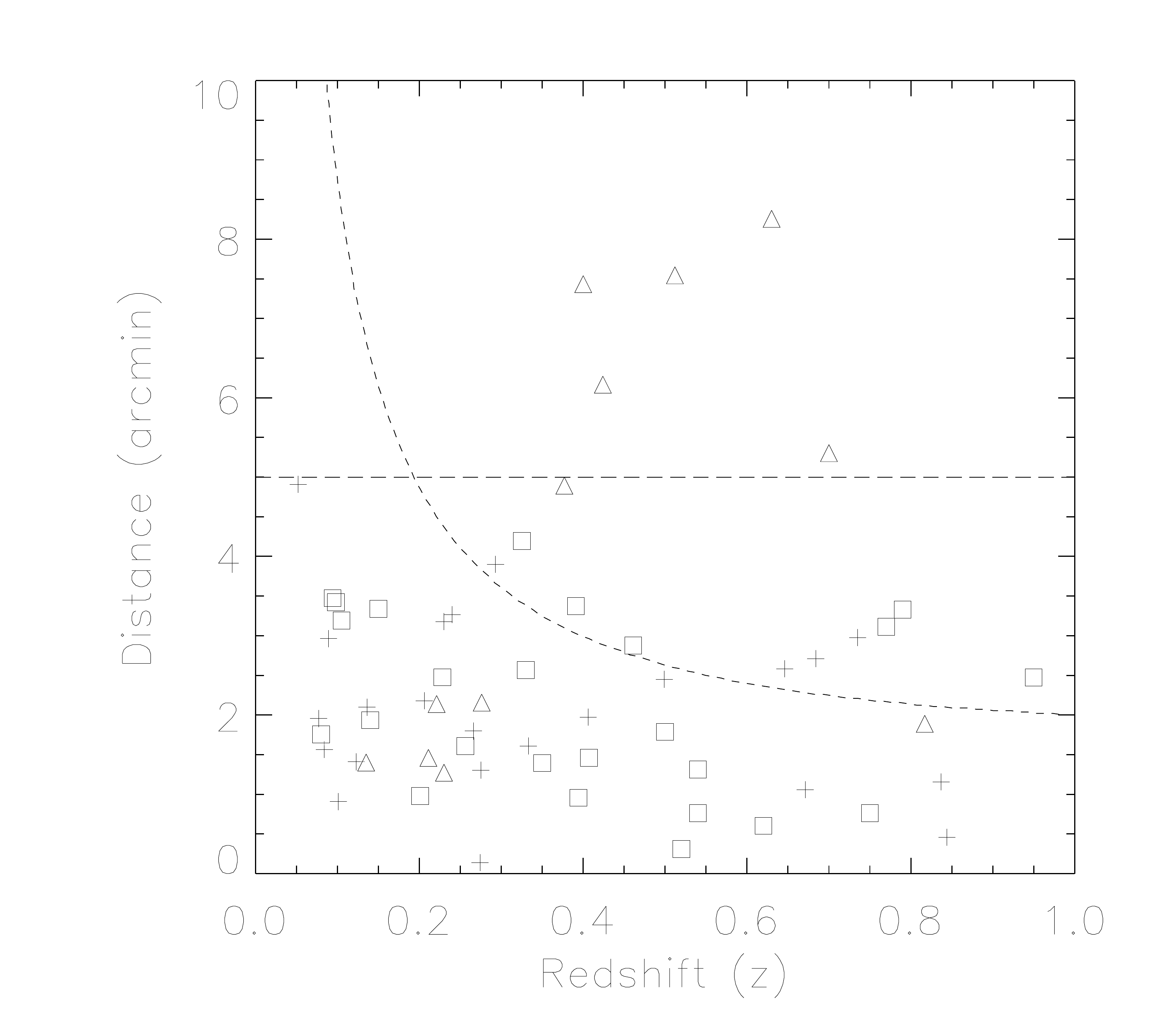}
\caption{Cluster optical centre offsets relative to their \Planck\ SZ position
as a function of cluster redshift. The dashed horizontal line at $5\arcm$ shows 
the maximum offset expected for a \Planck\ SZ detection (i.e.\ a \Planck\ beam). 
The dotted line corresponds to the angle subtended by 1\,Mpc in projection at 
the corresponding redshift. Symbols used are the same as in Figure \ref{fig:offsets}.}
\label{fig:d-vs-z}
\end{figure}

Figure~\ref{fig:d-vs-z} shows a similar analysis, where we plot the relative distance 
between the optical and SZ centres as a function of cluster redshift. Clusters at 
low redshift ($z<0.1$) present larger apparent radii, so, at this redshift, a 
typical virial radius of 1\,Mpc extends to about $10\arcm$. This means that SZ--optical associations at $5-10 \arcm$ 
distance could be expected. However, we find no optical counterparts at large 
distances. Even in the cases of clusters at $z<0.1$ the SZ--optical distance is below 
$5\arcm$.

\begin{figure}[ht!]
\centering
\includegraphics[width=\columnwidth]{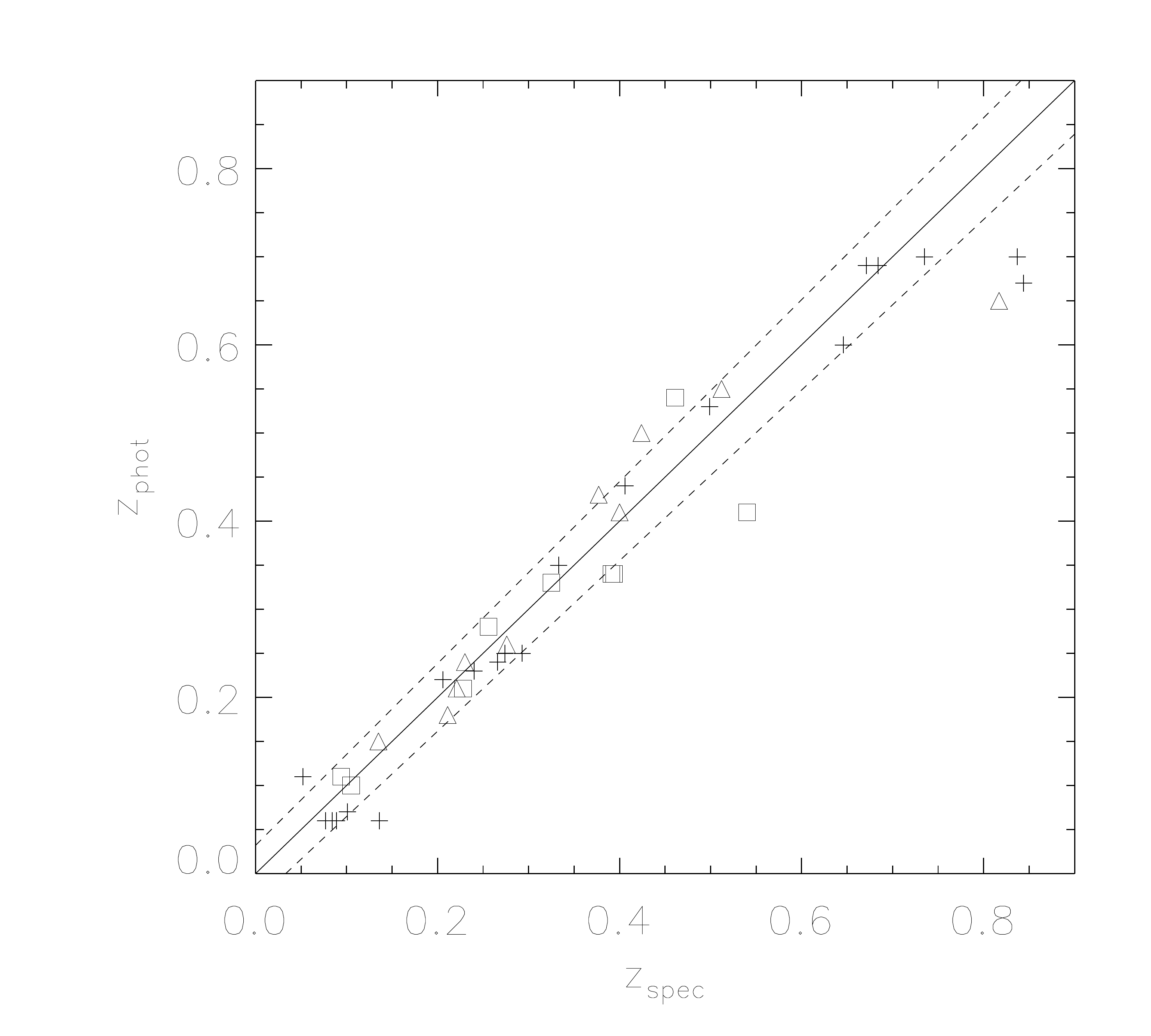}
\caption{Comparison between photometric estimates and spectroscopic redshifts.
The continuum line represents the 1:1 relation and dashed lines show the 
photometric redshift error $\delta z/(1+z) = 0.03$ when comparing clusters 
with $z<0.75$. Symbols used are the same as in Figure \ref{fig:offsets}.}
\label{fig:zphot}
\end{figure}

A comparison between photometric and spectroscopic redshifts for the clusters listed
in Table~\ref{tab:inpsz1} is shown in Figure~\ref{fig:zphot}. Photometric redshifts are 
estimated as described in Section~\ref{sec:photoz}. Clusters with $z>0.75$ are
excluded from this analysis because at this redshift range even the $r'-i'$ colour is
unusable to determine reliable photometric redshifts. For clusters at 
$z>0.75$ we would need photometry in the $z'$-band or even redder filters, which were not 
obtained in our imaging survey. Therefore, for clusters at $z_{\rm spec}>0.75$, the 
photometric redshift is expected to be lower than the actual value (see 
Fig.~\ref{fig:zphot}). This study yields a photometric redshift error of 
$\delta z/(1+z) = 0.03$ when comparing clusters with $z<0.75$.

\subsection{Notes on individual clusters}
\label{sec:notes}

In this section we present a detailed description of all the clusters in 
Table \ref{tab:inpsz1} that show particular features, such as multiple detection targets, 
substructural evidence of non-relaxation, fossil systems, or massive 
clusters with strong lensing effects. The SZ counterparts not discussed in this 
section seem to be regular clusters that do not present any difficulty in their 
association with the corresponding SZ signal or any peculiarity from 
the optical point of view. We also compare here identifications with other 
optical confirmations of \Planck\ PSZ1 sources. We note that all sky images presented 
in this section are orientated with north up and east to the left. 

\begin{figure}[ht!]
\centering \includegraphics[width=\columnwidth,keepaspectratio]{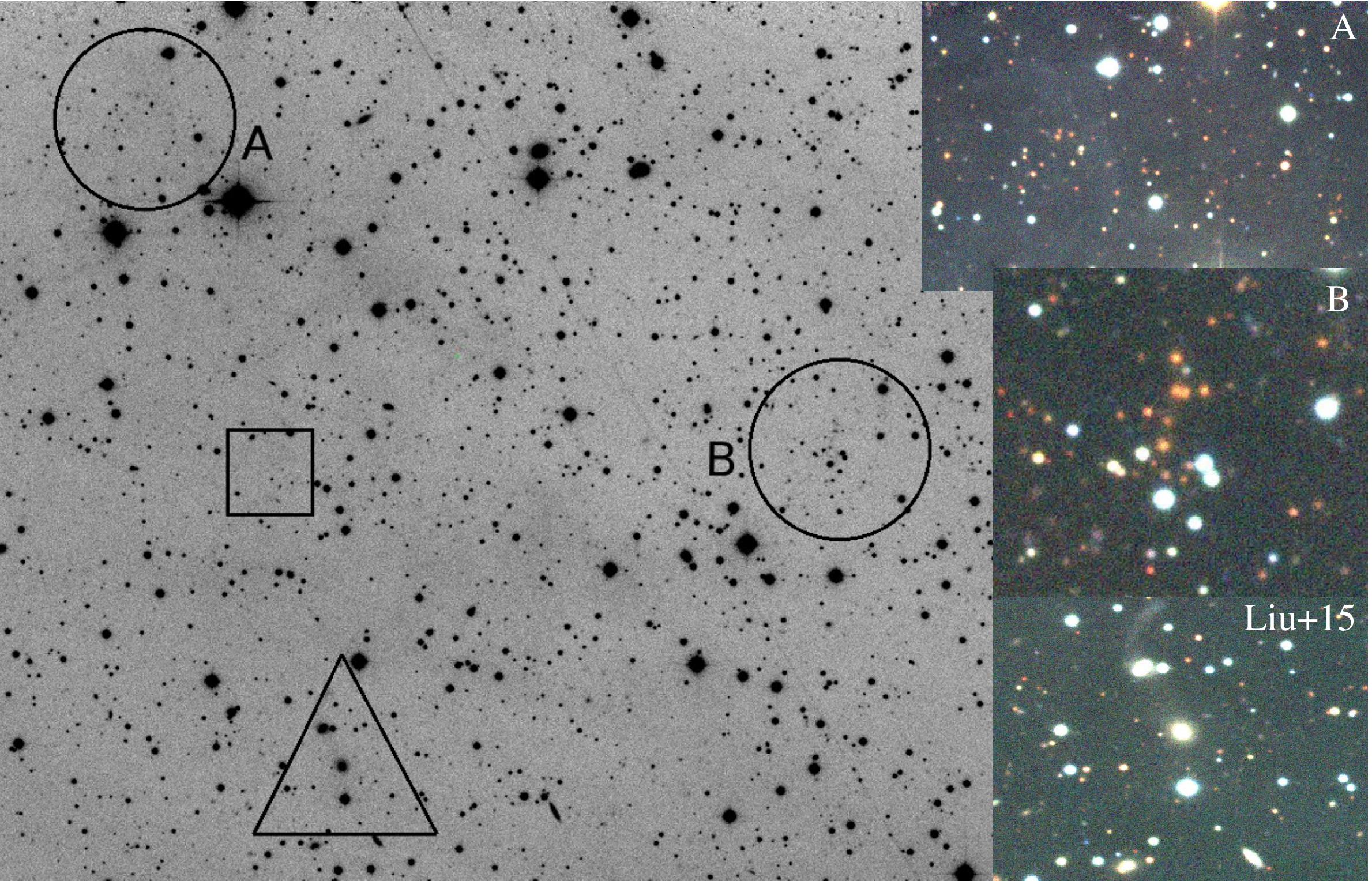}
\caption{Image obtained in the $r$-band with the WFC/INT for the PSZ1 G028.01$+$25.46
source covering a field of view of $10\arcm \times 10\arcm$. Circles A and B 
mark the respective positions identified in this article of these clumps at $z=0.658$ and $z_{phot}=0.60$. The square points to the \Planck\ PSZ1 
coordinates, and the triangle is located on the Liu+15 counterpart for this 
PSZ1 source. RGB colour images show zoomed regions in each clump. Note that, 
from our photometric analysis, we do not detect any important concentration of 
galaxies with $z_{\rm phot}=0.284$ around the Liu+15 coordinates.}
\label{fig:psz1-69} 
\end{figure}

\paragraph{PSZ1 G028.01$+$25.46} We identify two possible counterparts, which we call clumps A and B, for this 
SZ source (see Fig.~\ref{fig:psz1-69}). The 
A clump seems to be the richest counterpart. We performed multi-object spectroscopy 
using OSIRIS/GTC around clump A, and we confirmed this cluster by selecting
9 cluster members at $z=0.658$, showing a $\sigma_{\rm v} \sim 900$\,km\,s$^{-1}$. Clump 
B also shows a concentration of galaxies (poorer than clump A) with 
$z_{\rm phot}=0.60$ but remains at $6\arcm$ from the \Planck\ pointing. Consequently, 
we classify clump B as {\tt Flag}=3. \citet{Liu2015} (hereafter Liu+15), using 
the Pan-STARRS Survey \citep{magnier2013}, report an optical counterpart at 
$z_{\rm phot}=0.284$ around RA=17:11:45 and Dec=+07:15:17. Nevertheless, we do not detect 
any important concentration of galaxies at this redshift around the reported location, except for
a few galaxies that may configure a small system. Clump A is
therefore the only reasonable (and massive) cluster associated with the PSZ1 G028.01$+$25.46 
source, which we classify as {\tt Flag}=1, despite its centre being $5\parcm2$ (slightly 
$>5\arcm$) from the PSZ1 coordinates.

\begin{figure}[ht!]
\centering \includegraphics[width=\columnwidth]{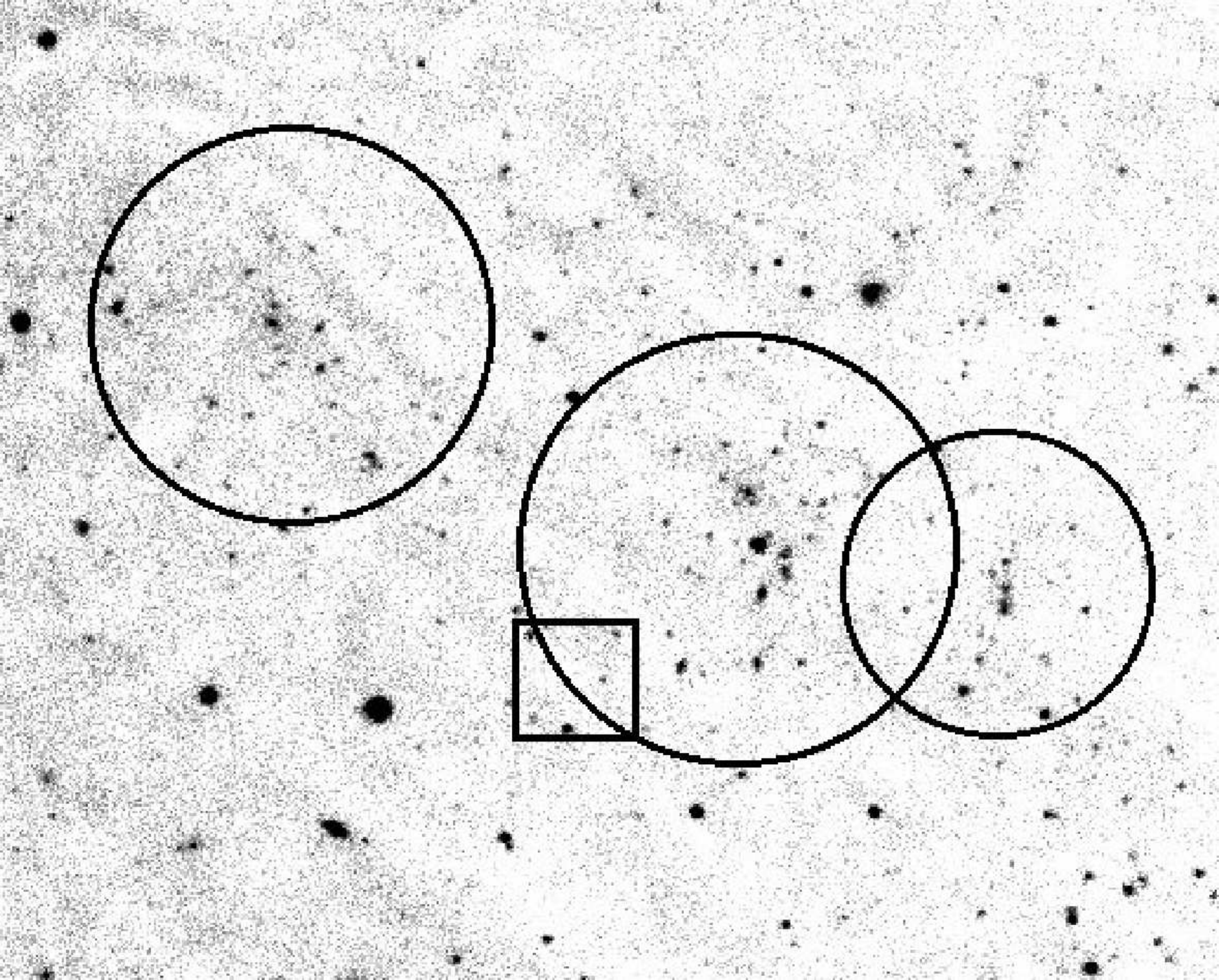}
\caption{$i$-band image, obtained with the WFC/INT, covering a $5\arcm \times 4\arcm$ FOV around 
the PSZ1 G108.26$+$48.66 cluster. Circles identify peaks in the galaxy density 
distribution consistent with $z=0.671$. The square represents the \Planck\ pointing 
of this SZ source.}
\label{fig:psz1-394} 
\end{figure}

\paragraph{PSZ1 G032.15$-$14.93, PSZ1 G042.96$+$19.11, PSZ1 G065.13$+$57.53, and PSZ1 G108.26$+$48.66} 
These systems are clusters at $z=0.377$, 0.499, 0.684, and 0.671 respectively. We confirm 
these systems photometrically and spectroscopically. We perform MOS using OSIRIS/GTC 
and we select 10, 8, 20, and 29 galaxy members for each system respectively. The analysis of 
the spatial distribution of likely cluster members, retrieved from WFC/INT data, shows evidence 
of the presence of substructures. PSZ1 G032.15$-$14.93 shows a peak around RA=19:43:11, Dec=-07:24:56, 
which represents the main body of the system, with  $\sigma_{\rm v} \sim 550$\,km\,s$^{-1}$, and a 
secondary peak, at $4\arcm$ towards the west, which corresponds to 1.2 Mpc at the redshift of the 
cluster. PSZ1 G042.96$+$19.11 and PSZ1 G065.13$+$57.53 also present a bimodal configuration. These 
objects were observed with ACAM/WHT. Analysis of the 2D galaxy distribution suggests a secondary 
substructure at 0.8 Mpc ($2 \parcm 3$) towards the south-east of the centre of the cluster, whereas the
PSZ1 G065.13$+$57.53 substructure is located at $5\arcm$ (2.7 Mpc) in the north-east with respect to 
the main body of the cluster. The global $\sigma_{\rm v}$ of this cluster seems to be about 720 and 
870\,km\,s$^{-1}$ respectively. PSZ1 G065.13$+$57.53 is one of the richest clusters found 
in this study. Finally, PSZ1 G108.26$+$48.66 is more complex, showing three peaks in the 2D 
galaxy density distribution. We identify a main body in the central position and two 
substructures almost aligned with the east--west direction (see Fig.~\ref{fig:psz1-394}). From the
radial velocities we estimate a global $\sigma_{\rm v} \sim 970$\,km\,s$^{-1}$. 

\paragraph{PSZ1 G044.92$-$31.66} Liu+15 validate this cluster as an actual counterpart. However,
after analysing our MOS data, obtained with DOLORES/TNG, we find only a very poor galaxy system
showing a low velocity dispersion ($\sigma_v < 500$ km/s). We therefore classify this system as
{\tt flag}=3, so that it is only very marginally associated with the SZ eource.

\paragraph{PSZ1 G075.29$+$26.66} This is a case of multiple detection. We select
two clusters within $6\arcm$ of the \Planck\ pointing. We map 
this region using WFC/INT and select two galaxy overdensities, the first
with a redshift of $z=0.282$ at $5\arcm$ towards the south of the \Planck\ pointing, 
and the second system with a estimated $z_{\rm phot}=0.30$ located at $5\arcm$ towards 
the NE. We perform MOS with DOLORES/TNG around RA=18:08:44.26 and Dec=+47:41:09
and we find 10 cluster members showing $\sigma_{\rm v} \sim 1000$\,km\,s$^{-1}$. 
Even though both clusters are farther than  $5\arcm$ from this PSZ1 target's coordinates,
we consider these two systems as actual counterparts of the SZ signal. In this
case, the SZ peak is located very close to the intermediate point between the two clusters
and, in this situation, the \Planck\ SZ signal could be a combination of the individual 
SZ effects produced by the two clusters. 

\paragraph{PSZ1 G081.56$+$31.03} This is an SZ target invalidated by vdB+16 owing to its low richness. 
The coordinates of the centre reported in Table \ref{tab:inpsz1} differ from those listed in vdB+16 by
less than $1\arcm$. The reason for this difference is probably that vdB+16 estimate the 
cluster centre as the location that maximizes the richness measurement, whereas our coordinates 
refer to the position of the brightest cluster member. This counterpart, classified as {\tt Flag}=2
by us, and that reported by vdB+16 correspond to the same galaxy distribution with photometric 
redshift estimates in agreement to within errors. Spectroscopic observations are needed in order to 
clarify whether this cluster is a massive system or not.

\paragraph{PSZ1 G090.48$+$46.89} We detected two systems around this SZ target.
First, we found a clump with $z_{\rm phot}=0.54$ at $4\arcm$ to the south of the 
\Planck\ pointing. This is probably the only system associated with this SZ target.
However, we detected a high-$z$ system ($z=0.676$) at a distance of $9\arcm$ to 
the east, by selecting 6 cluster members using OSIRIS/GTC MOS around  RA=15:44:07 
and Dec=+57:46:43.2, which seems to be too far from the \Planck\ SZ coordinates.

\begin{figure}[ht!]
\centering \includegraphics[width=\columnwidth,keepaspectratio]{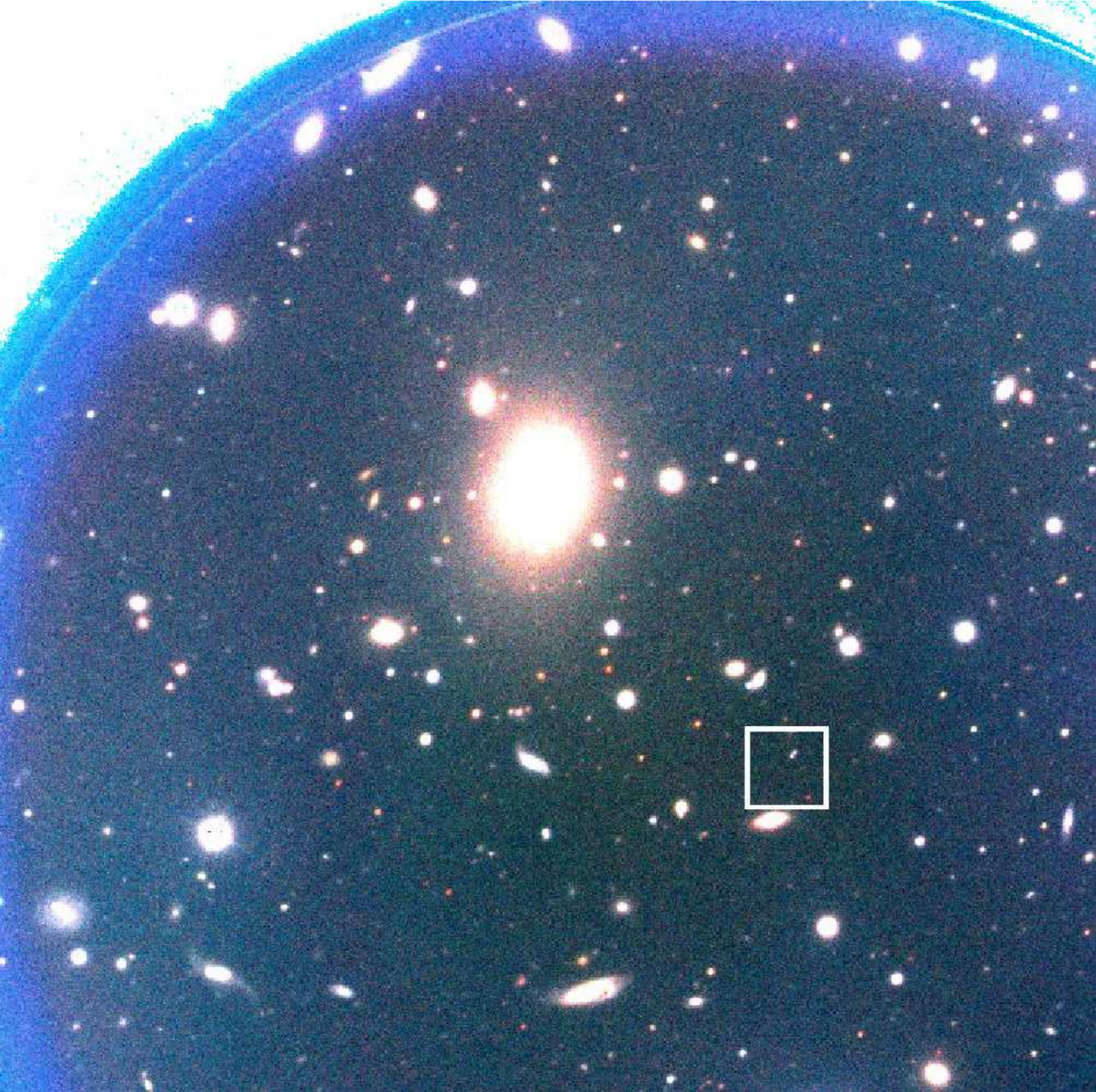}
\caption{Colour omposite RGB image performed using $g$-, $r$-, and $i$-band images
of the cluster PSZ1 G103.94$+$25.81 covering an area of $5\parcm5 \times 5\parcm5$
FOV. The huge BCG is clearly visible at $2\arcm$ to the NE of the \Planck\ PSZ1 
coordinates (square).}
\label{fig:psz1-374} 
\end{figure}

\paragraph{PSZ1 G103.94$+$25.81 and PSZ1 G185.93$-$31.21} These are two clear fossil 
galaxy systems. ACAM/WHT images (see Fig.~\ref{fig:psz1-374}) reveal huge BCGs at 
$2\arcm$ and $4\arcm$ to the north-east of their respective \Planck\ PSZ1 
coordinates. We confirmed these systems spectroscopically by selecting 17 and 2 cluster 
members at $z=0.077$ and 0.094 respectively. PSZ1 G103.94$+$25.81 shows  
$\sigma_{\rm v}\sim650$\,km\,s$^{-1}$.

\paragraph{PSZ1 G125.54$-$56.25} This is also a very complex region. We detected two
clusters at $2\arcm$ to the west of the \Planck\ PSZ1 coordinates, labelled
ID-458a and ID-458b in Table~\ref{tab:inpsz1}. These two systems almost 
overlap on the sky plane and their BCGs present $z=0.5475$ and $z=0.1676$. In 
addition, we selected a third system, ID-458c, at $3\parcm4$ to the north of the 
\Planck\ pointing. The three clusters present similar richness and all of them are
placed at $<5\arcm$ from the PSZ1 coordinates. They therefore probably all contribute
to the SZ signal. The 2D galaxy distribution and photometric redshift have been 
derived from our ACAM/WHT images, and spectroscopic redshifts were obtained 
from the SDSS DR12 database.

\paragraph{PSZ1 G142.1+157$+$37.28} In this case, we found a discrepancy with 
Liu+15, who select a cluster counterpart for this SZ source with $z=0.28$ 
located $6\parcm3$ to the SE of the \Planck\ PSZ1 coordinates. However, we 
detected a different system located at only $3\arcm$ to the north of the \Planck\ pointing. 
With the MOS observations using DOLORES/TNG we confirmed this cluster by selecting 6 galaxy 
members at $z=0.240$. From these few radial velocities, we derived a relatively high 
velocity dispersion, $\sigma_{\rm v}>700$\,km\,s$^{-1}$, which indicates 
that this system is massive and associated with the \Planck\ SZ signal. It 
remains unclear whether the ID-508 cluster of Liu+15 can also be considered an actual 
counterpart of this SZ source. However, another possibility is that both clusters are 
gravitationally bound.

\begin{figure}[ht!]
\centering \includegraphics[width=\columnwidth,keepaspectratio]{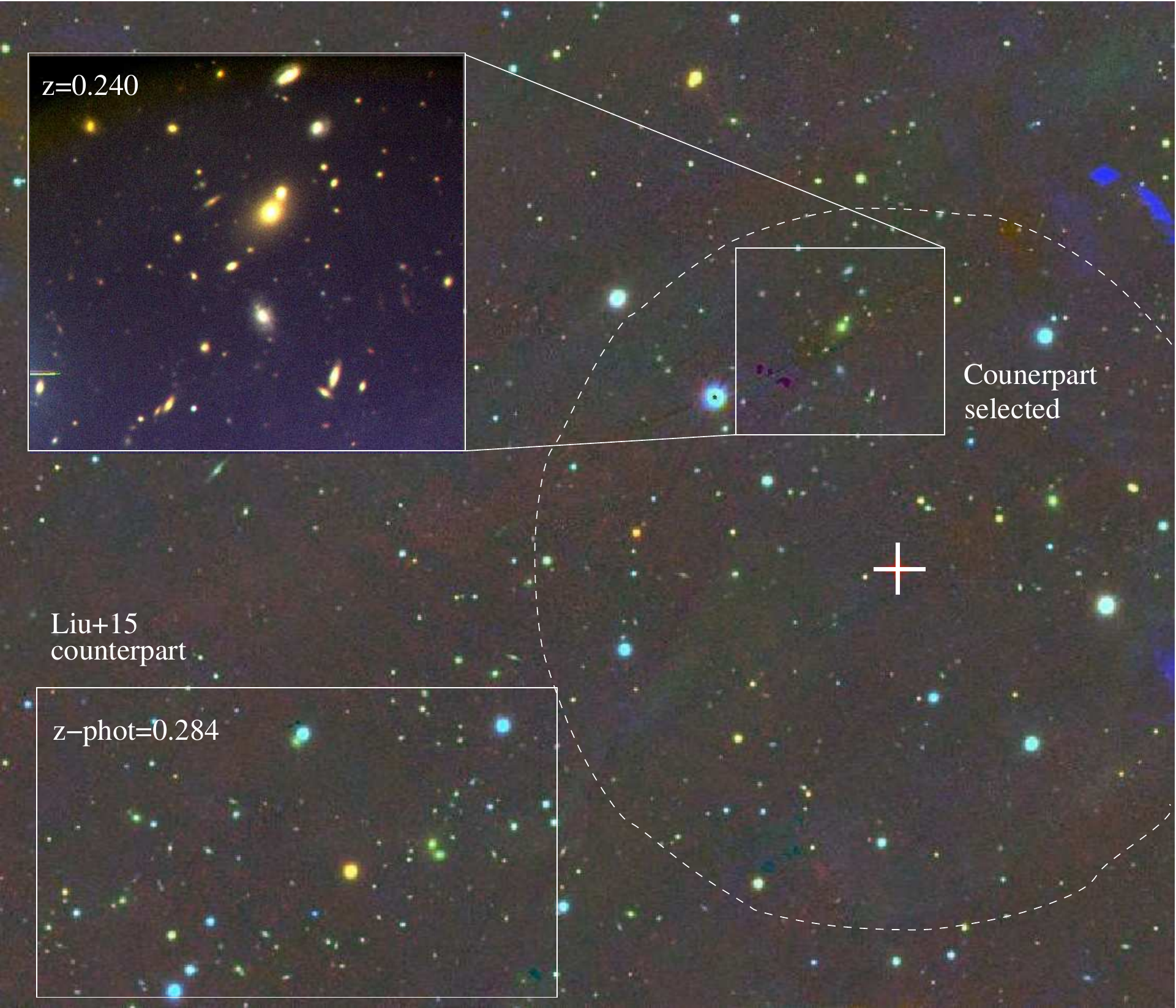}
\caption{PanSTARRS colour RGB image (composed of $g^\prime$, $r^\prime$, 
and $i^\prime$ filters) covering a $15\arcm\times13\arcm$ FOV around the PSZ1 G142.1+157$+$37.28 source. The cross and the dashed 
line circle correspond to the PSZ1 nominal pointing and $5\arcm$ selection area respectively. 
The cluster counterpart proposed in this study (at $z=0.240$) and that proposed by Liu+15 
(at $z_{\rm phot}=0.284$) are marked by solid rectangles. Superimposed is an ACAM/WHT zoomed 
image of the northern cluster is shown.}
\label{fig:psz1-508} 
\end{figure}

\paragraph{PSZ1 G151.80$-$48.06, PSZ1 G157.44$+$30.34 and PSZ1 G188.36$-$35.00} 
These three targets are known clusters, catalogued in \citet{planck2013-p05a},
\citet{planck2013-p05a-addendum}, and \citet{abell89} identified by correlating
\Planck\ PSZ1 sources with other known cluster catalogues. The three targets are 
ACO 307 \citep{abell89}, RX J0748.6$+$5940 \citep{Appenzeller1998}, and 
ZwCl 0401.8$+$0219 \citep{zwicky1961} respectively. However, their spectroscopic 
redshifts have been so far unknown. For that reason we obtained long-slit spectroscopy 
of the brightest galaxies of these systems. The two clear BCGs of ACO 307 present 
$z_{\rm spec}=0.1999$ and 0.2011. For both RX J0748.6$+$5940 and ZwCl 
0401.8$+$0219 we obtain 4 radial velocities of cluster members for each system, 
resulting in mean spectroscopic redshifts of $z=0.407$ and 0.275 respectively.

\paragraph{PSZ1 G169.80$+$26.10} This is a complex region showing multiple counterparts. 
We detected two well separated systems at high redshift ($z_{\rm phot}=0.73$ and
0.83 for A and B systems respectively) within $4\parcm2$ 
of the \Planck\ pointing. We observed this region using WFT/INT adding the $z^\prime$ 
band to the $g^\prime$, $r^\prime$, and $i^\prime$ Sloan filters. This allowed us to make a better
estimate of the photometric redshift, in particular for these high redshift clusters. 
Spectroscopy is needed in order to determine the precise redshift and $\sigma_v$ of
this system. However, the richness estimates agree with actual galaxy clusters. Thus,
given that these two systems are enclosed in a small projected region on the sky and very 
close to the PSZ1 coordinates, they are probably both contributing to the SZ signal 
of this PSZ1 source. 

\paragraph{PSZ1 G183.26$+$12.25} We also classify this target as a multiple detection 
source. Two systems are identified at $z_{\rm phot}=0.27$ and $z=0.638$ (see Fig.~\ref{fig:cmd}). 
Both systems present similar richness and are inside the $5\arcm$ region of the 
\Planck\ pointing, so the SZ signal of this PSZ1 source is probably a combination of the
individual SZ effects produced by these two clusters. More spectroscopic observations
are needed in order to estimate $\sigma_v$.

\begin{figure}[ht!]
\centering \includegraphics[width=\columnwidth,keepaspectratio]{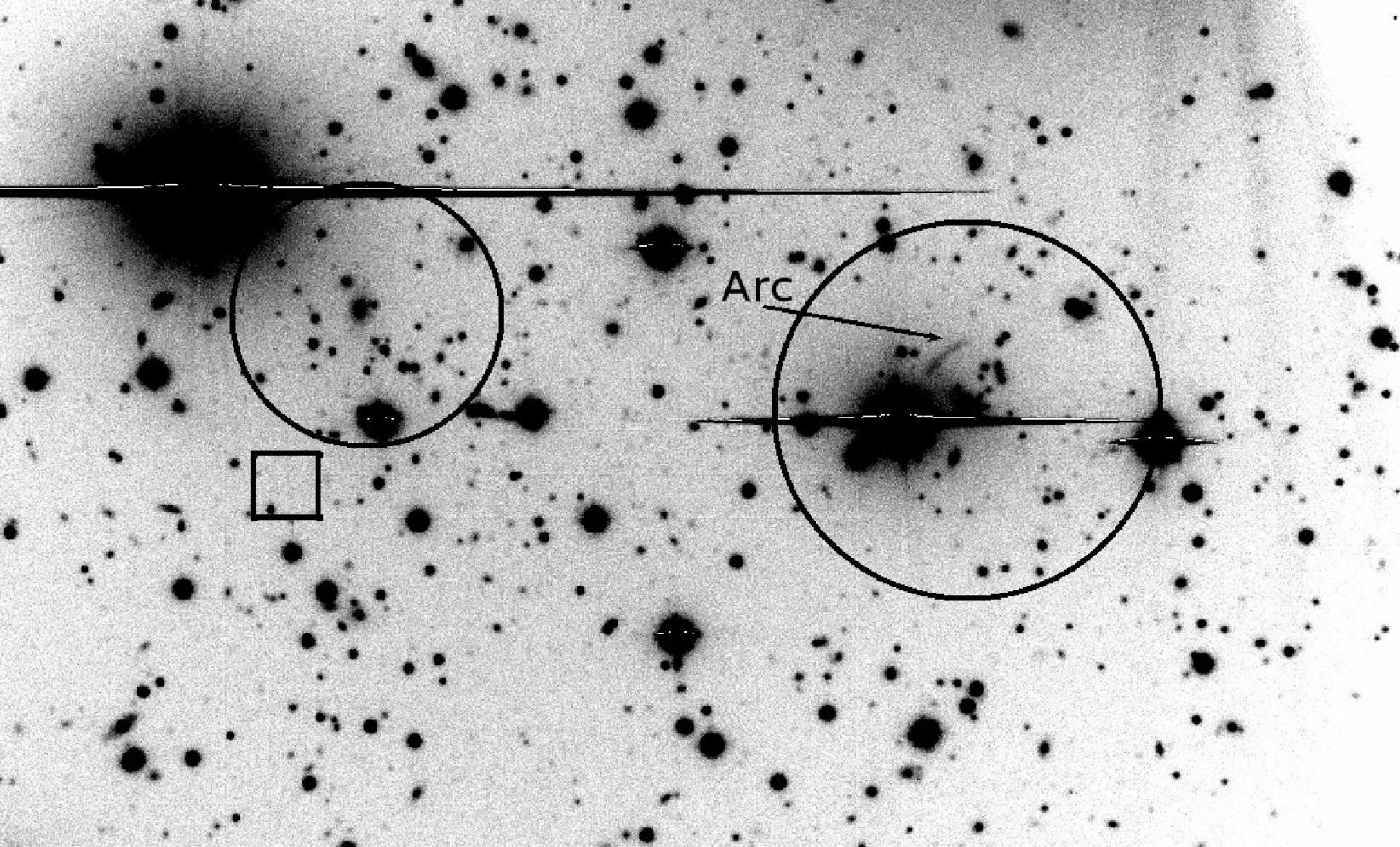}
\caption{$3\arcm\times5\arcm$ $r$-band image of the cluster PSZ1 G206.45$+$13.89 obtained 
with ACAM/WHT. A large gravitational arc, marked with an arrow in the plot, is clearly 
visible around the BCG. Circles mark the core of the main body of the cluster (to 
the west) and the substructure, detected in the 2D spatial distribution of galaxies and 
radial velocities. The box indicates the \Planck\ PSZ1 coordinates of this SZ target.}
\label{fig:psz1-681} 
\end{figure}

\paragraph{PSZ1 G206.45$+$13.89} This is the only cluster in our sample with strong 
lensing effects. We detect a significant gravitational arc around the BCG that is clearly 
visible in the cluster core (see Fig.~\ref{fig:psz1-681}). We performed MOS using 
OSIRIS/GTC and select 45 cluster members. The BCG is at $z_{\rm BCG}=0.4055$, and the 
global $\sigma_{\rm v}$ is $\sim 1200$\,km\,s$^{-1}$, which shows that this is a 
massive cluster. Moreover, the 2D galaxy distribution and the velocity field of this 
cluster show evidence of dynamical non-relaxation and the presence of substructure. 
The subcluster coincides very well with the \Planck\ PSZ1 coordinates, and the 
main body of the cluster is $2\parcm3$ to the west of the \Planck\ pointing.

\paragraph{PSZ1 G341.69$+$50.66} This is a cluster selected by Liu+15 as
a multiple detection. We obtained deep imaging with ACAM/WHT and performed
MOS observations with DOLORES/TNG, retrieving more than 70 redshifts of galaxies in
this field. Analysis of the velocity field of the galaxies reveals only one cluster
at $z=0.293$. We identified 31 galaxy cluster members and a global velocity dispersion 
of $\sim 800$\,km\,s$^{-1}$, showing clear evidence of dynamical non-relaxation. The 
main body, which contains the brightest galaxies and the densest galaxy distribution, 
is located at $4\arcm$ to the north-west of the \Planck\ PSZ1 coordinates. The secondary 
substructure is placed at RA=14:25:34.54 and Dec=-05:00:34.3, which corresponds to a 
distance of 2 Mpc from the main body at the redshift of the cluster. So, the \Planck\ 
PSZ1 target is at the mid-point bewteen the main body and substructure positions,
but we did not identify multiple counterparts in this SZ source.

\subsection{Non-detections and comparison with other previously studied counterparts}
\label{sec:nondetections}

The validation criteria described in Sect. \ref{sec:photoz} yield 62 SZ sources classified
as `non-detections' (ND) or as clusters weakly associated with the SZ signal detected by
\Planck\ \negthinspace. This means that no cluster counterparts were detected for 
54\% of the sample, where our deep imaging and spectroscopic data did not show any evidence 
of optical counterparts to the SZ sources. There are two possible ways to explain these 
non-identifications. The first, and most plausible, explanation is that there is no optical 
counterpart, owing to false SZ detections, high noise in the $Y_{500}$ \Planck\ maps 
(\citet{planck2013-p15}; see Fig. 4), or contamination in SZ maps produced by radio 
emission of galactic dust clouds. A second explanation is that the cluster counterpart 
does exist but is at high redshift ($z>0.85$), hence making it very difficult to detect at 
visible wavelength. 

In addition, We detected a few discrepancies with respect to previously reported counterpart validations. Whereas, overall, we fully agree with VdB+16 validation, Liu+15 validate some clusters as 
actual counterparts that we classify as ND and {\tt Flag}=3. The reason behind these 
disagreements is probably the different methods and restrictions imposed to validate
counterparts. 

In the following, we discuss the nature of some non-confirmed clusters in order to contextualize 
this kind of classification ({\tt Flag}=3 and ND).

\paragraph{PSZ1 G031.41$+$28.75} vdB+16 studied this SZ source as a cluster with 
$z_{\rm phot}=0.42$, a richness of $R=20.9\pm7.4$, and a Richness Mass\footnote{The 
Richness Mass is the mass of the cluster estimated from the mass--richness relation of 
\citet{rykoff2014} and \citet{rozo2015}.} of M$=1.07\pm0.4 \ 10^{14}$ M$_{\odot}$, located 
at $4\arcm$ towards the west of the \Planck\ pointing. However, vdB+16 invalidate this 
counterpart owing to its low richness mass. In agreement with vdB+16, we do not detect any 
significant galaxy concentration within a region of $5\arcm$ radius from the \Planck\ 
pointing, consistent with this photo-z. However, we detect a poor galaxy overdensity 
almost aligned with the PSZ1 coordinates and showing a coherent colour distribution in agreement 
with $z_{\rm phot}=0.24$. Long-slit spectroscopy observations, using OSIRIS/GTC, for 4 galaxy 
members confirmed this galaxy system to have a redshift of $z=0.230$, but more spectroscopic
observations are needed in order estimate its $\sigma_v$ so as to consolidate its validation 
as actual counterpart. For the moment, we classify this cluster as a very marginal counterpart 
({\tt Flag}=3) owing to its low richness.
 
\paragraph{PSZ1 G078.39$+$46.13} This is an SZ target studied by Liu+15, using SDSS DR12 
data, we also identify this cluster at $<z_{\rm spec}>=0.400$ and located at $7\parcm5$ to the 
south of the \Planck pointing. Liu+15 validated this cluster as an actual counterpart. However, 
following our validation criteria, we found this cluster to be too far from the PSZ1 coordinate to be 
considered a realistic counterpart. We therefore classified this cluster with {\tt flag}=3, indicating a weak 
association with this SZ target. No additional galaxy overdensities were found within the region 
of $5\arcm$ radius.

\paragraph{PSZ1 G135.92$+$76.21} We have performed MOS using OSIRIS/GTC and DOLORES/TNG and
retrieved more than 80 redshifts in this region. Among this spectroscopic sample, we 
detect two systems. One of them at $z_{\rm spec}=0.406$ and a second at $z_{\rm spec}=0.767$
(around RA=12:35:24.10, Dec=+40:30:08.02; these have not been included in Table \ref{tab:inpsz1} 
as a multiple detection). We detected 9 galaxy members in the former, and 11 members in 
the latter. However, the dynamical analysis reveals a $\sigma_v<500$\,km\,s$^{-1}$ and 
the optical richness is also R$<15$ in both two cases. So, despite these two systems'
possibly configuring actual galaxy groups, they are too poor and non-massive to be
associated with the SZ emission.

\begin{figure}[ht!]
\centering \includegraphics[width=\columnwidth,keepaspectratio]{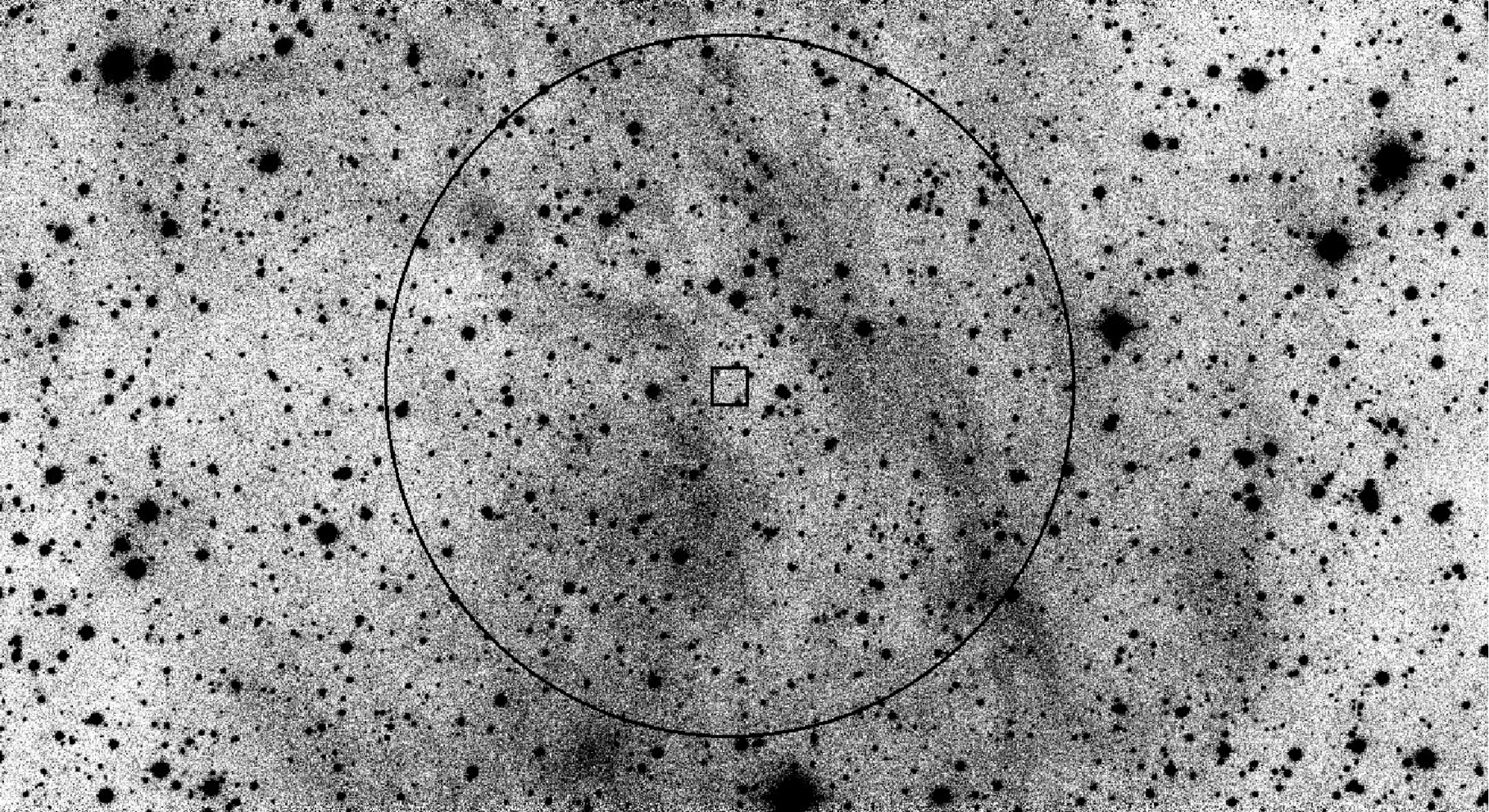}
\caption{WFC/INT $g$-band image of the target PSZ1 G001.00$+$25.71. The square 
marks the PSZ1 position, and the circle encloses the $5\arcm$ region around
the \Planck\ pointing. No cluster counterpart is identified in this case. The presence 
of significant galactic gas and dust structures may have influenced the SZ detection 
in this area.}
\label{fig:cirrus} 
\end{figure}

Galactic cirrus is common in targets of low galactic latitude. In fact, we 
detect large dust and gas structures in the images of the majority of our targets. Moreover, 
the presence of gas and dust galactic filaments is even very common in pointings at 
$|b|>25\deg$ but close to the galactic centre ($300\deg <l<60\deg$; see Fig.~\ref{fig:cirrus}). 
Half of the sample (57 clusters) are targets with $|b|<25\deg$, of which 32 
are classified as ND and only 3 with {\tt Flag}=3. These numbers reveal that about 
61\% of them are unconfirmed SZ targets located close to the galactic plane.

On the other hand, taking into account targets with $|b|>25\deg$, we classify 27 (47\%) 
sources as not confirmed (ND+`3'), while we detect 31 (53\% of confirmed clusters) 
of positive (`1'+`2') optical counterparts. Therefore, even though the optical follow-up  
presented here is incomplete, the numbers are indicative of the low purity 
of PSZ1 catalogue for targets with low SZ significance. In addition, we see a higher 
fraction of unconfirmed sources at $|b|<25\deg$, which suggests that 
targets located close to the galactic plane are more likely to present dust contamination
or even high SZ detection noise.

\paragraph{PSZ1 G053.50$+$09.56} This is a low galactic latitude field and is therefore
crowded with stars. This makes the detection of faint galaxies in the background 
very difficult. vdB+16 select this cluster, located at $4\arcm$ to the south of the PSZ1 
coordinates at $z_{\rm phot}=0.12$ with very low richness mass, M$\sim10^{13}$ M$_{\odot}$.
vdB+16 consequently discount this cluster as realistic SZ counterpart. In agreement with this result,
we do not detect any galaxy concentration with consistent photo-z in this field, we 
classify this target as a non-detection (ND).

\paragraph{PSZ1 G052.93$+$10.44 and PSZ1 G286.25$+$62.68} These are two special
cases. Photometric data obtained with WFC/INT show the presence of galaxy densities
well aligned with the \Planck\ pointing with $z_{\rm phot}=0.24$ and 0.18 
respectively. In fact, after obtaining DOLORES/TNG MOS data, we confirmed these clumps
as actual systems, retrieving 10 and 11 galaxy members for each cluster respectively.
However, the radial velocities obtained show a low $\sigma_{\rm v}$. Both structures
present $\sigma_{\rm v} < 350$\,km\,s$^{-1}$. So, although PSZ1 G052.93$+$10.44 
and PSZ1 G286.25$+$62.68 present optical counterparts, the corresponding galaxy systems
are probably very poor, being more group-like structures and hence very unlikely to be able to generate 
a significant SZ signal in \Planck\ maps. These cases illustrate how chance alignments
of poor clusters with SZ targets may introduce false identifications in SZ counts and how 
spectroscopy is the only way for revealing the actual nature of clusters by clarifying whether mass 
and $\sigma_{\rm v}$ take reasonable values to generate a detectable SZ signal for \Planck\ 
instruments.

\paragraph{PSZ1 G050.01$-$16.88, PSZ1 G098.67$-$07.03, and PSZ1 G345.81$+$42.38} 
We inspected these SZ targets using WFC/INT and ACAM/WHT deep images and  detected a few not very prominent
galaxy concentrations, but with a coherent $z_{\rm phot}$ 
and resembling galaxy clusters. However, after performing OSIRIS/GTC and DOLORES/TNG 
spectroscopy and analysing more than 40 radial velocities of galaxies in each
field, we did not select any actual cluster within these regions. The images of 
PSZ1 G050.01$-$16.88 and PSZ1 G098.67$-$07.03, as low galactic latitude sources, 
reveal the presence of diffuse galactic dust and gas.

\paragraph{PSZ1 G071.64$-$42.76} This PSZ1 source was studied by vdB+16, who invalidated 
this counterpart. This source shows the highest SZ significance ($S/N=8.82$) of all the targets included 
in this study. In agreement with vdB+16, the WFC/INT images do not show any significant galaxy 
overdensity. Only a very poor group of galaxies at RA=22:30:50.00 and Dec=+05 39 16.72 
compatible with $z_{\rm phot}=0.69$ is found. This region shows evidence of galactic dust 
contamination that is clearly visible in the $g$-band images. We therefore classify this PSZ1 target as 
ND. This case illustrates how the PSZ1 catalogue is not completely pure and how even high $S/N$ SZ 
sources may be false SZ detections.

A few cases are very representative of how SZ detection with high SZ significance can
be associated with low mass galaxy clusters. For example, PSZ1 G108.90$-$52.04, PSZ1 G157.44$+$30.34, 
and PSZ1 G148.20$+$23.49 reveal SZ $S/N>6.8$ but with R$<40$. On the contrary, 
PSZ1 G206.45$+$13.89 and PSZ1 G224.82$+$13.62 show very high optical richness (R$>120$) 
but moderate SZ $S/N$ ($<6$). Optical richness, R, and SZ significance do not necessarily have
to be correlated. This non-correlation can be explained in terms of the presence of galactic dust whose contribution contaminates the \Planck\ SZ detection, the Eddington bias in this sample,
and noise inhomogeneities in the \Planck\ maps. 

\paragraph{PSZ1 G090.14$-$49.71} Liu+15 validate this SZ source with a cluster at 
$z_{\rm phot}=0.207$ located $13\parcm2$ from the PSZ1 coordinates. Following our
selection criteria, we found no significant galaxy concentration in the ACAM/WHT images
showing coherent $z_{\rm phot}$ within $5\arcm$ from the PSZ1 position. Thus we classified this 
SZ target as ND.

\section{Conclusions}
\label{sec:conclusions}

This article is a continuation paper of the \Planck\ validation catalogue of SZ sources 
(PSZ1) published in \citet{planck2016-XXXVI}. The work presented here shows the results
of the first year of observations of the ITP13B-15A at the ORM, using the INT, TNG, WHT and 
GTC, as part of the optical follow-up validation and characterization programme 
of SZ \Planck\ sources.

We studied 115 SZ sources based on deep imaging and spectroscopy, which allowed us to 
analyse, in an unbiased way, the nature of systems based on their 2D galaxy distribution,
and radial velocities. By using photometric and spectroscopic observations we adopted a 
robust counterpart classification based on the alignment with respect to the \Planck\ SZ coordinates, 
the optical richness estimations, and the velocity dispersion of the clusters. We classified 
clusters as actual counterparts if they were well aligned with the SZ source, optically rich, and 
presented high $\sigma_v$ ({\tt Flag}=1). Also, clusters showing high R and good alignment, 
but with no MOS observations yet available, were classified as potentially confirmed clusters
with {\tt Flag}=2. Clusters poorly associated with the corresponding SZ source characterized 
by too low $\sigma_v$ or bad alignment were classified with {\tt Flag}=3, and non-detections
are regions with no galaxy concentration detected. Clusters flagged as `3' and ND
make up the unconfirmed cluster sample.

Following this classification, we found 53 confirmed counterparts, which means that 
46\% of this PSZ1 subsample were validated and characterized using optical data.
Sixty-two SZ sources remain unconfirmed. The cluster confirmations include 56 spectroscopic 
redshift determinations retrieved using long-slit and MOS observations. 
Table~\ref{tab:inpsz1} contains 6 multiple detections, for which a projection effect 
of two or more clusters can be associated with the SZ signal. We found two fossil systems, 
one cluster showing strong lensing effects and several systems with evidence of 
substructure. New spectroscopic redshifts are supplied for ACO 307, [ATZ98] B100, and 
ZwCl 0401.8+0219 galaxy clusters.

We found evidence for the effect of galactic dust contamination in the SZ detections. 
Galactic cirrus is present in the optical images, mainly in regions 
of low galactic latitude. We found that 61\% of SZ sources at low galactic latitudes 
($|b|<25\deg$) remained unconfirmed; this fraction is a bit lower (47\%) for SZ 
sources at $|b|>25\deg$. These numbers suggest that the presence of galactic dust and gas 
can lead to spurious SZ signals. Moreover, galactic dust contamination (together with 
other effects, such as Eddington bias and noise inhomogeneities in the  \Planck\ SZ detections) 
may distort the SZ $S/N$ estimation. The optical characterization, based on photometric and 
spectroscopic observations such us the work presented here represents an example of the 
need to use these kinds of techniques to validate and characterize SZ surveys.

The optical follow-up observations have recently terminated at the ORM, using the 
facilites described in this paper. We are currently analysing new \Planck\ PSZ1 and PSZ2 
data, and the results will be published in the near future.

\begin{acknowledgements}
This article is based on observations made with a) the Gran Telescopio Canarias
 operated by the Instituto de Astrof\'{\i}sica de Canarias, b) the Isaac 
Newton Telescope, and the William Herschel Telescope operated by the Isaac Newton 
Group of Telescopes, and c) the Italian Telescopio Nazionale Galileo  operated 
by the Fundaci\'on Galileo Galilei of the INAF (Istituto Nazionale di Astrofisica).
All these facilities are located at the Spanish 
del Roque de los Muchachos Observatory of the Instituto de Astrof\'{\i}sica de 
Canarias on the island of La Palma.
This research has been carried out with telescope time awarded by the CCI
International Time Programme at the Canary Islands Observatories (programmes
ITP13B-15A).
Funding for the Sloan Digital Sky Survey (SDSS) has been provided by the Alfred
P. Sloan Foundation, the Participating Institutions, the National Aeronautics
and Space Administration, the National Science Foundation, the U.S. Department
of Energy, the Japanese Monbukagakusho, and the Max Planck Society.
AAB, AF, AS, RB, DT, RGS, and JARM acknowledge financial support from Spain's 
Ministry of Economy and Competitiveness (MINECO) under the AYA2014-60438-P and 
ESP2013-48362-C2-1-P projects. HL is supported by the project PUT1627 of Estonian 
Research Council and by the project TK133, financed by the European Union through 
the European Regional Development Fund. JD, MA, and RFJvdB acknowledge support 
from the European Research Council under FP7 grant number 340519.
\end{acknowledgements}

\bibliographystyle{aa}

\end{document}